\documentclass[fleqn,usenatbib]{mnras}

\usepackage{blindtext}

\usepackage[T1]{fontenc} 
\usepackage{newtxtext,newtxmath}
\usepackage[dvipsnames]{xcolor}
\usepackage{gensymb}
\usepackage{multirow}
\usepackage{bm}
\usepackage{verbatim}
\usepackage{amsmath}
\usepackage{siunitx}
\usepackage{graphicx}
\usepackage{hyperref}
\usepackage{cleveref}
\newcommand{\be}{\begin{equation}}
\newcommand{\ee}{\end{equation}}
\newcommand{\nn}{\nonumber}
\newcommand{\bea}{\begin{eqnarray}}
\newcommand{\eea}{\end{eqnarray}}
\renewcommand{\d}{{\rm d}}
\newcommand{\ylm}[1]{Y_{\ell m}(\hat #1)}
\newcommand{\alm}{a_{\ell m}}

\DeclareMathOperator\erf{erf}
\DeclareRobustCommand{\VAN}[3]{#2}
\let\VANthebibliography\thebibliography
\def\thebibliography{\DeclareRobustCommand{\VAN}[3]{##3}\VANthebibliography}

\everymath{\displaystyle}

\title{Gravitational waves and galaxies cross-correlations: a forecast on GW biases for future detectors}
\author[S. Zazzera et al.]{
Stefano Zazzera,$^{1}$\thanks{E-mail: s.zazzera@qmul.ac.uk}
José Fonseca,$^{2,3}$
Tessa Baker$^{4}$
and Chris Clarkson$^{1,3}$
\\
$^{1}$Department of Physics \& Astronomy, Queen Mary University of London, Mile End Road, London E1 4NS, United Kingdom\\
$^{2}$Instituto de Astrofisica e Ci\^{e}ncias do Espa\c{c}o, Universidade do Porto CAUP, Rua das Estrelas, PT4150-762 Porto, Portugal\\
$^{3}$Department of Physics \& Astronomy, University of the Western Cape, Cape Town 7535, South Africa\\
$^{4}$Institute of Cosmology and Gravitation, University of Portsmouth, Burnaby Road, Portsmouth PO1 3FX, UK
}

\date{Accepted XXX. Received YYY; in original form ZZZ}

\pubyear{\the\year{}}

\begin{document}
\label{firstpage}
\pagerange{\pageref{firstpage}--\pageref{lastpage}}
\maketitle

\begin{abstract}
Gravitational waves (GWs) have rapidly become important cosmological probes since their first detection in 2015. As the number of detected events continues to rise, upcoming instruments like the Einstein Telescope (ET) and Cosmic Explorer (CE) will observe millions of compact binary (CB) mergers. These detections, coupled with galaxy surveys by instruments such as DESI, Euclid, and the Vera Rubin Observatory, will provide unique information on the large-scale structure of the universe by cross-correlating GWs with the distribution of galaxies which host them. In this paper, we focus on how these cross-correlations constrain the clustering bias of GWs emitted by the coalescence of binary black holes (BBH). This parameter links BBHs to the underlying dark matter distribution, hence informing us how they populate galaxies. Using a multi-tracer approach, we forecast the precision of these measurements under different survey combinations. Our results indicate that current GW detectors 
will have limited precision, with measurement errors as high as $\sim50\%$. However, third-generation detectors like ET, when cross-correlated with LSST data, can improve clustering bias measurements to within $2.5\%$. Furthermore, we demonstrate that these cross-correlations can enable a percent-level measurement of the magnification lensing effect on GWs. Despite this, there is a degeneracy between magnification and evolution biases, which hinders the precision of both. This degeneracy is most effectively addressed by assuming knowledge of one bias or targeting an optimal redshift range of $1 < z < 2.5$. Our analysis opens new avenues for studying the distribution of BBHs and testing the nature of gravity through large-scale structure.
\end{abstract}


\section{Introduction}\label{sec:intro}

Since the first detection of gravitational waves (GWs) in 2015 \citep{GWfirst}, they were rapidly established as powerful cosmological probes \citep{Mastrogiovanni_24}, for instance constraining the cosmic expansion history \citep{Abbott_2023}. The previous observing LIGO-Virgo-KAGRA (LVK)\citep{AdvLIGO,AdvVirgo,advKagra} runs have detected just short of 100 events \citep{ LIGOScientific,GWTC2, GWTC3_LVK}, however O$4$ is predicted to observe another $\mathcal{O}(100)$ GWs in its $2$-year period \citep{KAGRA:2013rdx}. The number of detections will grow with each observing run, such as O$5$, until third-generation ground based detectors will be completed. Instruments such as the Einstein Telescope (ET) \citep{Sathyaprakash_2010, Sathyaprakash_2012, Punturo_2010} and the Cosmic Explorer (CE) \citep{Evans_2021, Reitze_2019, Mpetha_2022} will provide us with an unprecedented amount of observations of binary black hole (BBH) mergers. Soon after their construction, which is planned to finish around 2035 \citep{Evans_2021,Maggiore_2020}, one expects the number of events to outpace the currently known number of merger events. For instance, ET is forecasted to detect around $10^6$ BBHs in its ten-year run \citep{Maggiore_2020}, a staggering four orders of magnitude higher than the number of BBHs known today. 

In a similar fashion, a suite of Stage IV dark energy experiments is now starting, both as ground-based telescopes as well as space missions. These include DESI \citep{DESIp1,DESIwp}, Euclid \citep{EuclidOG,Euclid_2019} and the Vera Rubin Observatory carrying out the Legacy Survey of Space and Time (LSST) \citep{LSST_2021,Sanchez_2022,LSST_book}. More futuristic concepts as Megamapper \citep{Astro2020,Megamapper} are being proposed, and will be due in the next decade. These will measure tens of millions of galaxy redshifts across a wide range of galaxy types; for instance, DESI will target bright galaxies (BGS) at low redshift, luminous red galaxies (LRG) at intermediate $z$, and bright emission lines galaxies (ELGs) at higher redshifts. 

Galaxies are well known, and well explored, tracers of the underlying dark matter distribution of the Universe \citep{Martinez_1999,Padmanabhan_2007,Bonvin_2008,Hui_2008,Chen_2008,Yoo_2009,Bonvin_2011,Challinor_2011,Jeong_2012,Tansella_2018,Castorina_2022}. However, they also host all the astrophysical processes which are thought to produce BBHs. Thus, BBHs can likewise be used as biased tracers of the dark matter. There is a further underlying assumption here, that black holes can only be formed astrophysically. In fact if BBHs are not solely astrophysical then the value of the clustering bias would be different from the one of a sole astrophysical origin. While we note this possibility we will not discuss these in this paper. 

Without an electromagnetic counterpart, a GW signal from a BBH merger carries distance information only in the form of a luminosity distance (once the redshifted chirp mass is constrained). Recent work by \citet{Us} has carried out the different clustering statistics required for objects living in luminosity distance ($D_L$) space as opposed to redshift space, highlighting the discrepancies especially at large scales. This meant constructing the number counts fluctuation in luminosity distance space, which was found to be dependent on three different tracer-dependent bias parameters: clustering, magnification and evolution bias \citep{Maartens_2021,Zazzera_2024}. Whilst the former describes the connection between the dark matter density fluctuation to the tracer-specific density fluctuation, the latter two describe effects caused by the limited sensitivity of our detectors in accessing our full past lightcone. In particular, the magnification bias accounts for objects being magnified in/out the detector's threshold, while the evolution bias shows how well we can trace the cosmic evolution of the chosen tracers given a detector.

Measuring these biases would inform us on several properties of the BBHs. Whilst magnification and evolution bias would constrain properties of the BBH population such as chirp mass distribution or intrinsic merger rate, measuring the clustering bias would provide a direct link to the dark matter distribution, acting as probe to differentiate between black holes tracing dark matter halos and black holes living in isolation, whether of primordial origin or else, such as in \citet{Scelfo_2020}.

One way of measuring these parameters is through cross-correlations, and in particular exploiting the powerful multi-tracer approach. Whilst cross-correlating GWs sources with other tracers has been explored to put constraints on cosmological parameters and test General Relativity \citep{Afroz:2024joi,Balaudo:2022znx,Mukherjee:2022afz,Mukherjee:2019wfw,Mukherjee:2019wcg,mukherjee2018,Mukherjee_2021,Cigarr_n_D_az_2022}, cross-correlating well-known tracers like galaxies with a more unconstrained one such as GWs has also been suggested as tool to investigate several properties of GWs \citep[e.g.][]{Libanore_2021,Libanore_2022,Scelfo_2018,Scelfo_2020,Scelfo_2022,Scelfo_2022_2}. Here we expand on previous studies in a number of ways. Firstly, we consider the differences in the number count fluctuation between redshift space (RS) and luminosity distance space (LDS) for the different tracers used, i.e. galaxies in RS and GWs in LDS (see \citet{Balaudo_2024} for thorough work on this as well). Secondly, we include all relativistic corrections to the number counts, and include effects due to both lensing magnification and the evolution of sources, i.e. magnification and evolution biases. Finally, we also examine different galaxy surveys to construct a timeline of forecasted measurement on the GWs clustering bias.

The paper is structured as follows. We start by describing the fundamental equations expressing clustering in LDS in  \autoref{sec:setup}. Subsequently, we describe the properties of each tracer and survey used in  \autoref{sec:app_survey}, and the Fisher formalism we adopted in  \autoref{sec:fisher}. In  \autoref{sec:clustering} we then present a forecast on the amplitude of the clustering bias of GWs using different combination of current and future surveys. Further, in  \autoref{sec:epsilons} we present the forecasted errors on, respectively, relativistic effects and on the magnification and evolution biases of GWs when using a ET$x$LSST-like combination of surveys. Finally, in  \autoref{sec:conclusion} we summarise and draw our conclusions.

\section{Number Count Fluctuation}\label{sec:setup}
In this section we summarise the underlying equations describing clustering in both luminosity distance and redshift space. We start by reminding the reader the definition of the number counts fluctuation (or density contrast):
\be\label{eq:density_contrast}
\Delta_N=\frac{N-\langle N\rangle}{\langle N\rangle}\,,
\ee
where $N$ is the number of objects seen in a given (observed) direction $\boldsymbol{n}$ at a given observed space (be it luminosity distance $D_L$ or redshift $z$), and $\langle N\rangle$ is an average over directions. Further, we define the line element used to derive the results below:
\be\label{eq:metric}
ds^2=a^2(\eta)\left[-(1+2\Psi)d\eta^2+(1-2\Phi)\delta_{ab}dx^a dx^b\right]\,,
\ee
where $\Phi$ and $\Psi$ are the metric potentials and $\eta$ is conformal time.

For a tracer living in redshift space such as galaxies, the observed density contrast in \autoref{eq:density_contrast} can be expanded into \citep{Bonvin_2008,Challinor_2011}:
\bea
\Delta^g(\boldsymbol{n},z) &=& b^g\delta_n-\frac1{\mathcal{H}}\partial_r(\bm v\cdot\bm n) \nn\\
&& +\frac{5s^g-2}{2\bar r}\int_0^{\bar r} \d r\ \frac{\bar{r}-r}{r}\Delta_{\Omega}(\Phi+\Psi)\nn\\
&&+\left[\frac{5s^g-2}{\bar{r}\mathcal{H}} -5s^g+b_e^g - \frac{\mathcal{H}'}{\mathcal{H}^2}\right](\bm v\cdot\bm n)\nn\\
&&+ g_{gal}(\Psi,\Phi,r) \, . \
\eea
The first term is the matter density contrast, the second the redshift-space distortions, the third the magnification lensing contribution, the fourth the Doppler term and $g_{gal}(\Psi,\Phi,r)$ is a function grouping further GR effects which are subdominant. For a full expression please refer to \citet{Bonvin_2008} and \citet{Challinor_2011}. The astrophysical parameters $b^g$, $s^g$ and $b_e^g$ are the clustering, the magnification and evolution biasses respectively, and depend on the particular galaxy catalog considered.

In the case of gravitational wave merger events without counter parts we live in luminosity distance space. Therefore we follow \citep{Us} and write the generic expression for the number counts as:
\bea\label{eq:coefficients}
\Delta^{GW}(\boldsymbol{n},D_L) &=& b^{GW}\delta_n+A_D(\bm v\cdot\bm n)+A_{LSD}\partial_r(\bm v\cdot\bm n) \nn\\
&& +\int_0^{\bar r} \d r\ \frac{A_{L}}{\bar r}\Delta_{\Omega}(\Phi+\Psi) + g_{GW}(\Psi,\Phi,r) \, . \
\eea
Here $\delta_n$ is the density contrast in the Newtonian gauge and $b^{GW}$ the clustering bias of the gravitational wave events. Here we only report the main correction terms, i.e. Doppler, Luminosity distance Space Distortions (LSD) - radial velocity distortions - and lensing, with amplitudes
\bea
&A_D& = 1-2(\gamma+\beta) \label{eq:AD}\, ,\\
&A_{LSD}& = -2\ \frac{\gamma}{\mathcal{H}} \label{eq:ALSD}\, ,\\
&A_{L}& = \frac12 \left[\left(\frac{\bar r-r}{r}\right)(\beta-2)+\frac{1}{1+\bar r \mathcal{H}}\right]\,\ \label{eq:AL} 
\eea
and we group the rest of the GR corrections under the last term $g_{GW}(\Psi,\Phi,r)$.
We note that $\bar{r}$ is the source's position, $r$ the integral's comoving distance. We define 
$\gamma \equiv \bar{r}\mathcal{H}/(1+\bar{r}\mathcal{H})$, and 
\be\label{eq:beta}
\beta\equiv
1-5s^{GW}+\gamma\left[\frac2{\bar r \mathcal{H}}+\gamma\left(\frac{\cal{H}'}{\mathcal{H}^2}-\frac1{\bar r\cal{H}}\right)-1-b_e^{GW}\right] \, ,
\ee
where $s^{GW}$ and $b_e^{GW}$ are the magnification and evolution biases, respectively. An in-depth analysis of these parameters for GWs is presented in \cite{Zazzera_2024}.

We expand the fields $\Delta(\boldsymbol{n},x)$ in a spherical harmonic decomposition 
\be
\Delta(\boldsymbol{n},x)= \sum^{\infty}_{\ell = 0} \sum^{\ell}_{m = -\ell} \alm(x) \ylm{n}\, ,
\ee
where $\ylm{n}$ are the spherical harmonics, and $x$ is a generic variable which can be the redshift $z$ or luminosity distance $D_L$. In this basis the $\alm$ preserve the statistical properties of $\Delta$. Therefore the first non-zero n-point function is the two-point function, the angular power spectrum:
\be
\langle \alm a^*_{\ell'm'} \rangle \equiv C_\ell\ \delta_{\ell\ell'}\delta_{mm'}\,.
\ee
This can then be computed fully, including the fact that the data is generally binned into intervals in distance, to yield the angular power spectrum for the $i$th and $j$th bins:
\be\label{eq:cl_gen}
C^{ij}_\ell= 4\pi \int {\rm d}\ln k\ \Delta^{A,i}_{\ell}(k)\Delta^{B,j}_{\ell}(k)\ {\cal P}_{\cal R}(k)\,,
\ee
where ${\cal P}_{\cal R}(k)$ is the power spectrum of primordial perturbations and $\cal R$ is the curvature perturbation. This is effectively cross-correlating counts of tracer $A$ in the $i$th bin with those of tracer $B$ in bin $j$. The functions $\Delta^{A,i}_{\ell}(k)$ take into account the radial window function $W$ and distribution of sources $\partial n^A/{\partial x}$, i.e., 
\be\label{eq:aps_ldist}
\Delta^{A,i}_{\ell}(k)=\int {\rm d}x\ W(x,x_i)\ \frac{\partial n^A}{\partial x}(x) \ \Delta^{A}_{\ell}(k,x) \,.
\ee

\section{Survey  properties}\label{sec:app_survey}

In this section we describe the properties of each survey used, for both GWs and galaxies. We include the number densities of target sources, together with expressions for the clustering, magnification and evolution biases. 

\subsection{GW experiments}
We assume all observed GWs are emitted by the same population of BBHs, with a merger rate following a standard Madau-Dickinson rate
\citep{YeFishback,MadauDickinson}:
\bea \label{eq:madau}
R_{GW}(z) = R_0\frac{(1+z)^{2.7}}{1+(\frac{1+z}{2.9})^{5.6}} \, ,\
\eea
with $R_0$ providing the merger rate at $z=0$ and is given by \citet{GWTC2,GWTC3_LVK} as $R_0=23.9$ Gpc$^{-3}$yr$^{-1}$. We note here that the values used for  \autoref{eq:madau} are not fully known and in this paper we fix them to follow a regular Madau-Dickinson model. This is due the fact that astrophysical BBHs are the results of stellar processes, thus it is common to assume they will too follow a rate similar to the stellar formation rate. We can then set the amplitude $R_0$ with LVK observations, and when 3G detectors will come online we will have better constraints on the rest of the parameters. A different merger rate will impact the angular power spectrum of GWs as it will result in different values of the magnification and evolution biases. Additionally, it will yield a new estimate of the shot noise. 

The number density of observed GW sources is then modelled following previous studies \citep{Zazzera_2024,Oguri_2018} as:
\begin{equation}\label{eq:GW_numdens}
    n_{\rm GW}(z) = \tau\frac{R_{GW}(z)}{1+z}\int \d \mathcal{M}\ \phi(\mathcal{M})\,S(\rho_{th};\mathcal{M},z) \,, 
\end{equation}
where $\tau$ is the observation time of the detector, $R_{GW}(z)$ is the intrinsic merger rate in \autoref{eq:madau}, $\phi(\mathcal{M})$ is the chirp mass $\mathcal{M}$ distribution, with
\bea \label{eq:chirp_def}
\mathcal{M} \equiv \frac{(m_1m_2)^{3/5}}{(m_1+m_2)^{1/5}} \, .\
\eea
We assume the primary mass $m_1$ to follow a \textit{Power-Law + Peak model} \citep{GWTC2,GWTC3_LVK}), i.e. a normalised power-law over the range $[5,85]M_\odot$, and with a Gaussian peak introduced to model a pile-up from pulsational pair-instability supernovae \citep{Talbot_2018}. The secondary mass $m_2$ is then given through a conditional mass ratio distribution, modelled as a power-law \citep{GWTC3_LVK}. The two distributions are then combined to obtain the chirp mass PDF following Appendix B of \citet{Zazzera_2024}.

For a full explanation of the survival function $S(\rho_{th};\mathcal{M},z)$ we refer the reader to \citet{Zazzera_2024, Oguri_2018, Finn_1996}. Notably, the function contains the power spectral density of the experiment, thus making each bias detector-dependent. 
Finally, we set the horizon redshift for each GW experiment as the redshift at which we are able to observe $<100$ sources (see \autoref{tab:sumspecs}). 

Following \citet{Zazzera_2024}, the magnification and evolution biases for each GWs experiment are defined as:
\bea
 s^{GW} &\equiv& -\frac{1}{5}\frac{\partial\log n_{GW}}{\partial\log\rho_{th}}\bigg|_{a} \, ,\\
 b_e^{GW} &\equiv& \frac{\partial\log n_{GW}}{\partial\log a}\bigg|_{\rho_{th}} \, ,
\eea
where $\rho_{th}$ is the signal-to-noise ratio (SNR) threshold of detection. This is used in lieu of the flux, commonly used in the galaxy counterpart. Additionally, the extra factor of $1/5$ is to fix the expression of the number counts (\autoref{eq:density_contrast} and \autoref{eq:beta}) and setting $s$ accordingly. This was done to keep the number counts in luminosity distance general. 

We them compute the biases from 
\autoref{eq:GW_numdens}, and fit 
a cubic spline in redshift $a+bz+cz^2+dz^3$ to parameterise the results in terms of $z$ and speed up computation. We report the coefficients of the polynomials in \autoref{tab:GWsmagbiases}. 


The two biases described so far have a strong impact on the number counts (and thus the angular power spectrum) as they are present in many correction terms. In particular, they both show in the Doppler and lensing terms, as shown in the factor $\beta$ in eq. \autoref{eq:beta}, which includes both $s$ and $b_e$. Not accounting for these biases has been shown to result in large discrepancies at higher redshift \citep{Zazzera_2024}. However, they are also rich with information regarding the population of BBHs, as they are built using both their merger rate and their chirp mass distribution.   

Finally, we address the clustering bias. This has been explored e.g. in \citet{Libanore_2021}, with a Halo Occupation Distribution method applied to a simulation, and where they found it could be described by a simple model, scale-independent and linearly dependent on redshift. A linear model was assumed also in \citet{Calore_2020}, and in fact it is commonly done so for galaxies as well, at least to first approximation \citet{Ruiz_Macias_2021, Euclid_2019, LSST_2021}. Therefore, we assume a simple two-parameter model of the form:
\bea\label{eq:clus_bias}
b = B (1+z)^\alpha \, ,
\eea
with $B$ and $\alpha$ assumed to be unity, for simplicity. Constraining such a toy model with future facilities, and its cross-correlations with galaxy surveys will be the focus of section \ref{sec:clustering}.


\subsection{Galaxy surveys}

Similarly to the GWs case, we first model the number density of observed galaxies. This reads, in general, as:
\bea\label{eq:gal_numdens}
n_{g}(z) &=& \int_{-\infty}^{M_{cut}}{\rm{d}}M\ \mathcal{F}(z,M) \, ,
\eea
where $\mathcal{F}(z,M)$ is the luminosity function of the galaxies considered as function of the absolute magnitude $M$. 
The magnification and evolution biases are then calculated following \citet{Maartens_2021}:
\bea
s &=& \frac{\partial\log_{10} n_g}{\partial M_c} \, ,\\
b_e &=& \frac{\partial\log n_g}{\partial\log a} \, .\
\eea

However, contrary to the GWs experiments considered earlier, each survey employed in this paper targets different galaxy types, resulting in different parameterisations of their number densities, and also of clustering, magnification and evolution biases. Whilst these parameters are all calculated similarly, some studies provided fitting formulae which both speed up and simplify the computations. We report all luminosity functions and fitting functions to number densities used in \autoref{sec:specs}.

We select four different surveys for the scope of this paper:
\begin{itemize}
    \item \textit{DESI BGS-like}:
    For a DESI-like survey we select only the Bright Galaxy Sample (BGS), in the redshift range $z\in[0,0.6]$ \citep{DESIp1,Ruiz_Macias_2021}, so as to overlap well with the redshift range of the current LVK observing run O$4$.
    \item \textit{Euclid-like spectroscopic}: spectroscopic survey targeting $H\alpha$ galaxies in the range $z\in[0.9,1.6]$ \citep{Euclid_2019}.
    \item \textit{LSST-like:}
    Photometric LSST-like survey, targeting galaxies at redshifts $0<z<3$ \citep{LSST_2012,LSST_2021,LSST_book}.
    \item \textit{Megamapper-like}:
    futuristic spectroscopic survey, targeting Lyman-break Galaxies (LBG) in the redshift range $2\lesssim z\lesssim 5$, following \citet{Astro2020,Sailer_2021,Megamapper}.
\end{itemize}
We summarise important specifications such as redshift range and sky coverage of each survey in \autoref{tab:sumspecs}.

\section{Fisher formalism}\label{sec:fisher}

In this section we describe the Fisher forecast formalism used in this paper. We define the data covariance as:
\bea\label{eq:gamma}
\Gamma^{ij}_{\ell} = C_{\ell}^{ij} + \mathcal{N}^{ij} \, ,
\eea
where $C_{\ell}^{ij}$ is the angular power spectrum of bins $i,j$ we which to observe 
(see \autoref{eq:cl_gen}), and $\mathcal{N}^{ij}$ is the corresponding noise power spectrum and is usually independent of the multipole $\ell$.

In a multi-tracer approach, using a GW survey and a galaxy survey, we can represent schematically the data covariance matrix as:
\bea\label{eq:multitracer_cov}
\Gamma_\ell (z_i,z_j) &=& \begin{bmatrix}
\Gamma_{\ell,ij}^{gal,gal} & \Gamma_{\ell,ij}^{gal,GW}\\
\Gamma_{\ell,ij}^{GW,gal} & \Gamma_{\ell,ij}^{GW,GW} \
\end{bmatrix} \, .
\eea

The noise angular power spectrum for a tracer is dominated by the shot noise:
\bea\label{eq:shot_noise}
\mathcal{N}^A_{i} = \frac1{N^A_i} \, ,
\eea
where $N^A$ is the number of objects (galaxies or GWs) per steradian in the $i$-th bin
\bea\label{eq:shot_noise2}
N^A_i = \int \d z \frac{\partial n^A}{\partial z} W(z,z_i;\Delta z_i,\sigma_i^z) \, ,
\eea
where $\partial n^A/\partial z$ is the comoving number density of objects as function of redshift, $W$ the window function centred at $z_i$, bin size $\Delta z_i$, and redshift scatter $\sigma^z_i = \sigma_0 (1+z)$. 

In the case of GWs, all quantities are commonly defined in terms of the luminosity distance, as it is the sole distance measure one has access from the data. Irrespective of this, once a fiducial cosmology is set, one can interchange between variables, i.e., between redshift $z$ and luminosity distance $D_L$ using the duality relation $D_L=(1+z)\int \d z'/H(z')$. Therefore the number density of observed sources is defined in terms of redshift (as in \autoref{eq:GW_numdens}) and we can then use the same expression as in \autoref{eq:shot_noise2} to compute the number of events in a data bin. Although the version of \texttt{CAMB} used in this work requires a redshift input for the construction of an angular power spectrum in luminosity distance, once we set a fiducial cosmology we keep consistency. In fact, both in redshift space and luminosity distance space, \texttt{CAMB} uses the fiducial cosmology to translate variables into background conformal time, variable in which it computes all quantities. For the purpose of the calculation of the Fisher matrix one uses a 5-point stencil numerical derivative which consistently changes the fiducial cosmology to gauge the dependence of the observable on a given cosmological parameter.

\begin{table}
    \centering
    \begin{tabular}{|c|c|c|c|c|c|c|}   
    \hline
    Survey & $z$-range & $\Delta z$ & $\sigma_z$ & $A_{survey}$ \\
    \hline
    O4 -like & $[0-1]$ & $0.4$ &$0.2$ & $4\pi$ \\
    O5 -like & $[0-1.4]$ & $0.4$ & $0.2$ & $4\pi$ \\
    ET -like & $[0-3.0]$ & $0.7$ & $0.1$ & $4\pi$ \\
    \hline
    DESI-like BGS & $[0-0.6]$ & $0.1$ & $0.02(1+z)$ & $15000 {\rm deg}^2$ \\
    Euclid-like & $[0.9-1.6]$ & $0.1$ & $0.001(1+z)$ & $14000 {\rm deg}^2$ \\
    LSST-like & $[0-3]$& $0.1$ & $0.02(1+z)$ & $18000 {\rm deg}^2$ \\
    Megamapper-like & $[2-5]$ & $0.1$ & $0.02(1+z)$ & $14000 {\rm deg}^2$ \\
    \hline
    \end{tabular}
    \caption{Summary of specifications of the surveys considered. We display the redshift range, size of $z$ bins used, i.e. $\Delta z$, the redshift scatter $\sigma_z$ and the area of the sky sampled.}
    \label{tab:sumspecs}
\end{table}

A list of the values of $\sigma_0$ is found in \autoref{tab:sumspecs} for each survey.
In the case of cross-correlating different tracers 
we set, for simplicity, the corresponding shot noise to zero, i.e., $\mathcal{N}^{ij}=\mathcal{N}^{i}\delta^{ij}$. Shot-noise comes from the correlation function at the same object. We do expect that some objects may overlap between a GW and galaxy catalog. One can model the cross-shot noise as proportional to the number density from the overlap in halo mass range of the two tracers weighted by the number densities of each tracer in consideration. Therefore we expect the overlap to be small, i.e., low cross-shot noise. For more details look at Appendix A of \citet{2020JCAP...09..054V}.
We understand that the results obtained will therefore be optimistic, although in the context of a simple Fisher analysis we consider this acceptable.

For a galaxy survey, the window function to define the binning in distance is given by a combination of error as \citep{Ma_2006,Viljoen_2021}:
\bea\label{eq:window}
W(z,z_i;\Delta z_i,\sigma_i^z)  = \frac1{2} \left[ \erf\left(\frac{z_i+\delta z_i -z}{\sqrt{2}\sigma_i^z} \right) -\erf\left(\frac{z_i-\delta z_i -z}{\sqrt{2}\sigma_i^z} \right) 
\right] \, .
\eea
The window function, similarly to the number of objects in the i-th bin (eq. \autoref{eq:shot_noise2}), is calculated in redshift for GWs as well. Ultimately however, there is no difference between integrating the expression in redshift, and integrating one in which $z$ is switched with $D_L$, as the number of sources will be the same.

Finally, to account for sky localisation uncertainty we apply to the angular power spectra of GWs a beam. We assume that to first order we can model this as Gaussian:
\bea\label{eq:beam}
B_\ell = \exp\left(-\frac{\ell(\ell+1)}{16\ln 2}\theta^2_{res}\right) \, ,
\eea
where we set the GWs resolution $\theta_{res} = 5\degree$ \citep{Libanore_2022}, consistent with distribution of localisation of 3G data \citep{Sathyaprakash_2012, Punturo_2010}. This effectively reduces the signal at smaller scales due to the limiting resolution of the detector.

\begin{figure*}
    \centering
    \includegraphics[width=\linewidth]{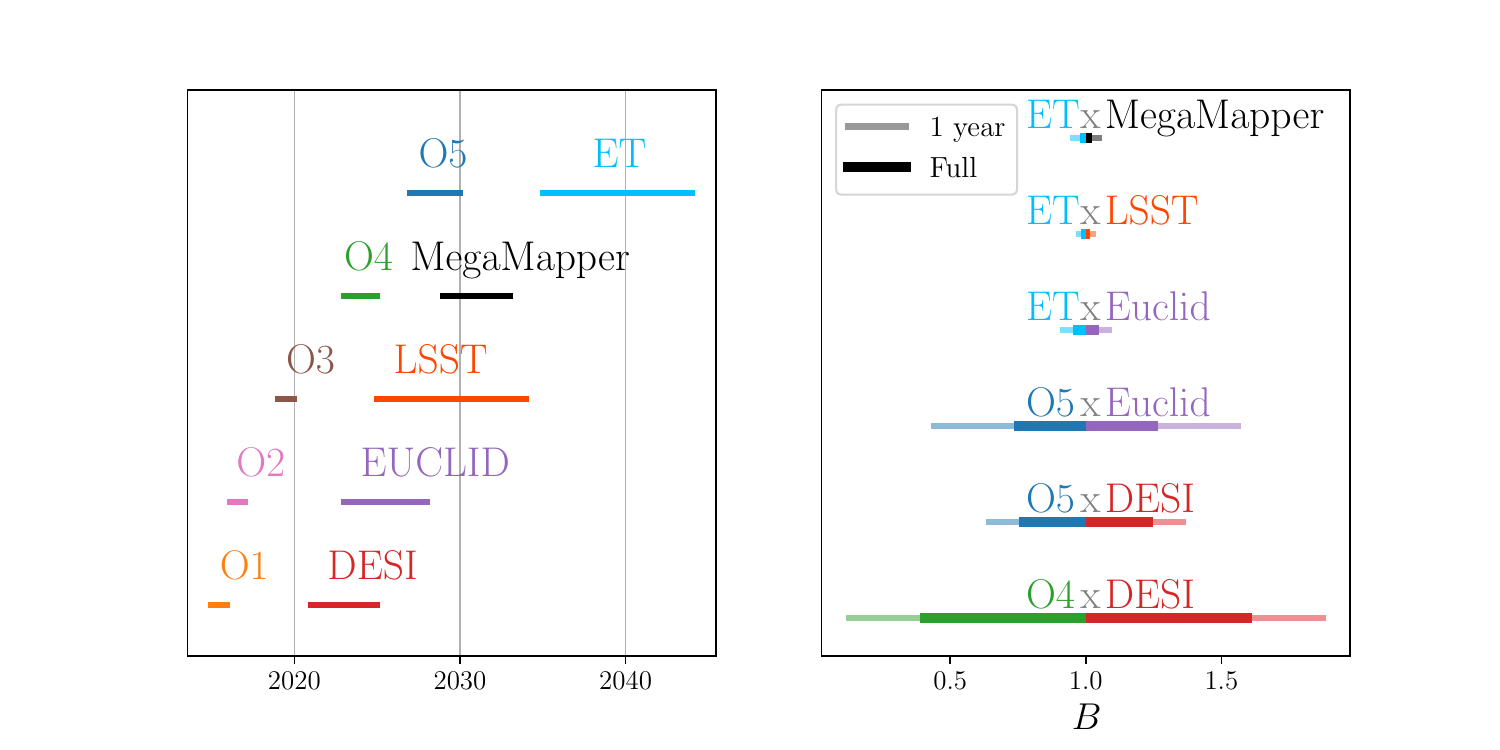}
    \caption{\textit{Left}: Timeline of GWs experiments and galaxy surveys considered in this paper. O$1-5$ stand for the different LVK observing runs. Vertical displacement is for clarity only. \textit{Right}: Measurement of the amplitude $B$ of the clustering bias of GW sources by cross-correlating different pairs of GW experiments and galaxy surveys. Faint lines indicate only $1$ year of observation, whilst solid ones include the (predicted) full length of the experiment. In the case of an ET-like and LSST-like survey we only adopt 5 years of observations to show the potential with half their predicted lifetime. Note that each detector and survey name should be accompanied by ``-like", e.g. O$4$-like x DESI-like. For stylistic reasons we leave it out to render the figure neater.}
    \label{fig:timeline}
\end{figure*}

The Fisher matrix is constructed following \citet{Abramo_2022}:
\be\label{eq:fisher_entries}
F^{\mu\nu}_{\bar{\ell}}=\frac{f_{sky}}{2}\displaystyle\sum_{\ell\in\bar{\ell}}(2\ell+1)\text{Tr} \left\{ \frac{\partial\Gamma_{\ell}^{\imath\jmath}}{\partial\theta^{\mu}}\left[\Gamma^{\jmath\imath\prime}_{\ell}\right]^{-1}\frac{\partial\Gamma_{\ell}^{\imath\prime\jmath\prime}}{\partial\theta^{\nu}}\left[\Gamma^{\jmath\prime\imath}_{\ell}\right]^{-1}\right\} ,
\ee
where $\theta$ is a parameter vector and $f_{sky}$ is the fraction of sky observed. For GW detectors we take $f_{sky}=1$ as they are sensitive to all directions. In reality, there would be a difference whether the source is seen from directly above/below the instrument, or from the sides. However, with a network of detectors around the globe this problem is greatly reduced. The sky area for galaxy surveys is found in \autoref{tab:sumspecs}.

Finally, our fisher matrix analyses we will always consider, and marginalise over, the standard cosmological parameters:
\bea
\theta_c = \{H_0, \Omega_b,\Omega_{cdm},A_s,n_s\}\, ,
\eea
set to the fiducial values given by Planck 2018 results \citep{Planck_2018}. We will also consider additional parameters in our analyses which we describe in the following sections.

\section{Forecasted clustering bias constraints}\label{sec:clustering}

We then proceed to construct a full multi-tracer Fisher forecast by computing the auto and cross angular power spectra of GWs and galaxies. As the former live in luminosity distance space, whilst the latter live in redshift space, we compute the $C_\ell$ with a modified  version of the publicly available code \texttt{CAMB} and presented in \citet{Us}. Therefore, we are able to construct the angular power spectra for such correlations accounting for the different expressions of the relativistic effects in the two spaces (see Table 1 in \citet{Us} for a full comparison). \citet{Us} showed that at large scales the difference between the two could reach $\sim20\%$.

We include magnification and evolution biases of GWs in this analysis. However, in this particular section of the paper we initially keep them as fixed parameters as opposed to marginalising over them, since they will only affect the relativistic corrections and not the density term, which is coupled to the clustering bias only.

Using \autoref{eq:gamma} to \autoref{eq:fisher_entries} we can thus construct cross-correlations between different GW experiments and galaxy surveys, and proceed to compute a Fisher forecast analysis with the aditional parameters $\theta \in \{B,\alpha,\theta_c\}$, thus assuming that all bias parameters of the galaxy samples considered - clustering, magnification and evolution - are known and fixed (thus not marginalising over them). We show on the left of \autoref{fig:timeline} the different observing runs of LVK and Einstein Telescope, together with four galaxy surveys (DESI BGS, Euclid, LSST and Megamapper), and their expected timeline of observation. 
\autoref{tab:sumspecs} summarises these. In particular, we display the binning size used for each survey to achieve optimal results. A larger bin size implies lower shot noise as more sources are encompassed, however it decreases localisation accuracy. We tried different bin sizes, and report the results only for the optimal cases. 

\begin{figure*}
    \centering
    \includegraphics[width=\linewidth]{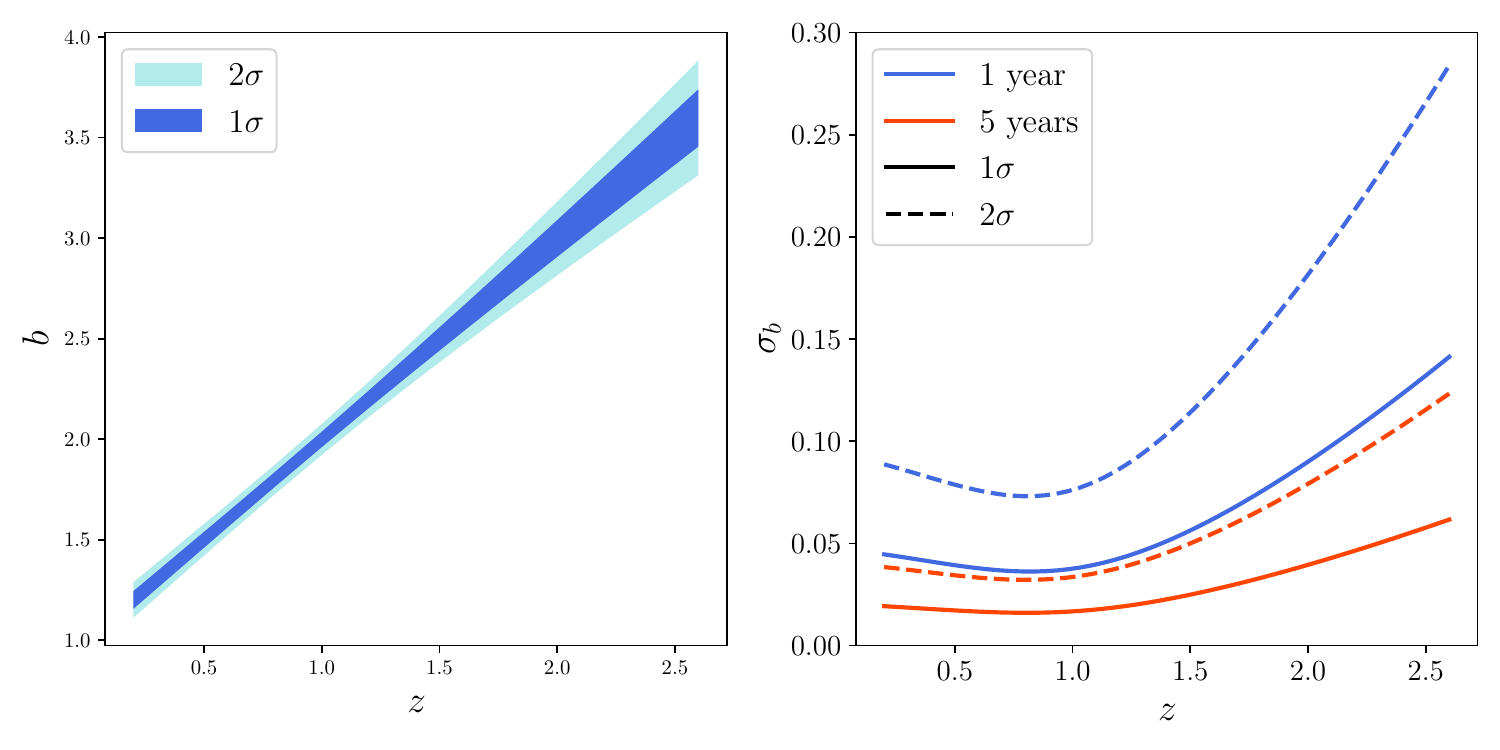}
    \caption{\textit{Left}: Value of the $1$ and $2\sigma$ contour on the clustering bias $b$ as a function of redshift for a $1$ year observation period using cross-correlations of GWs from an ET-like experiment and galaxies from an LSST-like survey. \textit{Right}: Forecasted $1$ and $2$ sigma contours on the clustering bias of GWs, for both $1$ and $5$ years of observations, respectively in blue and red lines.}
    \label{fig:clus_bias_ratio}
\end{figure*} 

We show the results on the right hand side of  \autoref{fig:timeline}, where we present the $1\sigma$ uncertainty on the amplitude of the clustering bias, i.e. the value of $B$ in  \autoref{eq:clus_bias}, found by cross-correlating a different combination of GWs experiments and galaxy surveys. We find that, whilst a combination of current experiments such as O$4$-like x DESI-like BGS yields a large error on the value of the bias, the precision largely increases with future detectors. In particular, with an O$5$-like experiment a precision of $\sim30\%$ is achieved when cross-correlating with a DESI-like BGS survey. It's also interesting how a better precision is obtained by such a combination, as opposed to cross-correlating with a EUCLID-like survey. This is due to the difference in redshift range covered by the two galaxy surveys: the former focuses on nearby redshifts, where shot noise for O$5$ is low, whereas the latter targets objects further away, resulting in a higher GW shot noise. Thus the overlapping redshift range is optimal only in the case of cross-correlating with the BGS sample. 

However, the exciting results lie in cross-correlations with third generation detectors, such as an ET-like experiment. In the best-yielding combination within the surveys considered, i.e. ET-like $\times$ LSST-like, we find a measurement error of $\sim2.5\%$ with $5$ years of observations. Percent-level precision is achieved with an ET $\times$ Megamapper-like cross-correlation, however with slightly higher error, likely because there are fewer redshift bins covered by both surveys.

Selecting ET $\times$ LSST, we proceed to examine the clustering bias as a function of redshift. 
We proceed by sampling the distribution of each parameter and evaluating the corresponding value of $b$. Further, we compute the mean and the standard deviation for each sample at every value of redshift. Finally, we plot the resulting one and $2\sigma$ errors as a function of redshift on the left hand side of \autoref{fig:clus_bias_ratio}. 

Additionally, we plot the values of the $1$ and $2\sigma$ error on the measurement of $b$ as a function of $z$ on the right hand side of \autoref{fig:clus_bias_ratio}. We do so for both $1$ year of observation and for $5$ years. Interestingly, we note that the uncertainty is lowest just before $z=1$. This is likely due to a lower shot noise induced by the higher number of sources predicted in our model of the merger rate, which is a simple Madau-Dickinson rate (see \autoref{sec:app_survey}).
We forecast a precision high enough to confidently measure the clustering bias of GWs even with just a single year of observations, at least up to $z<1.5$, as the $1-$sigma error is less than $5\%$.

\section{Relativistic Effects}\label{sec:epsilons}

\subsection{Measurement of GR effects}

\begin{figure*}
    \centering
    \includegraphics[width=\linewidth]{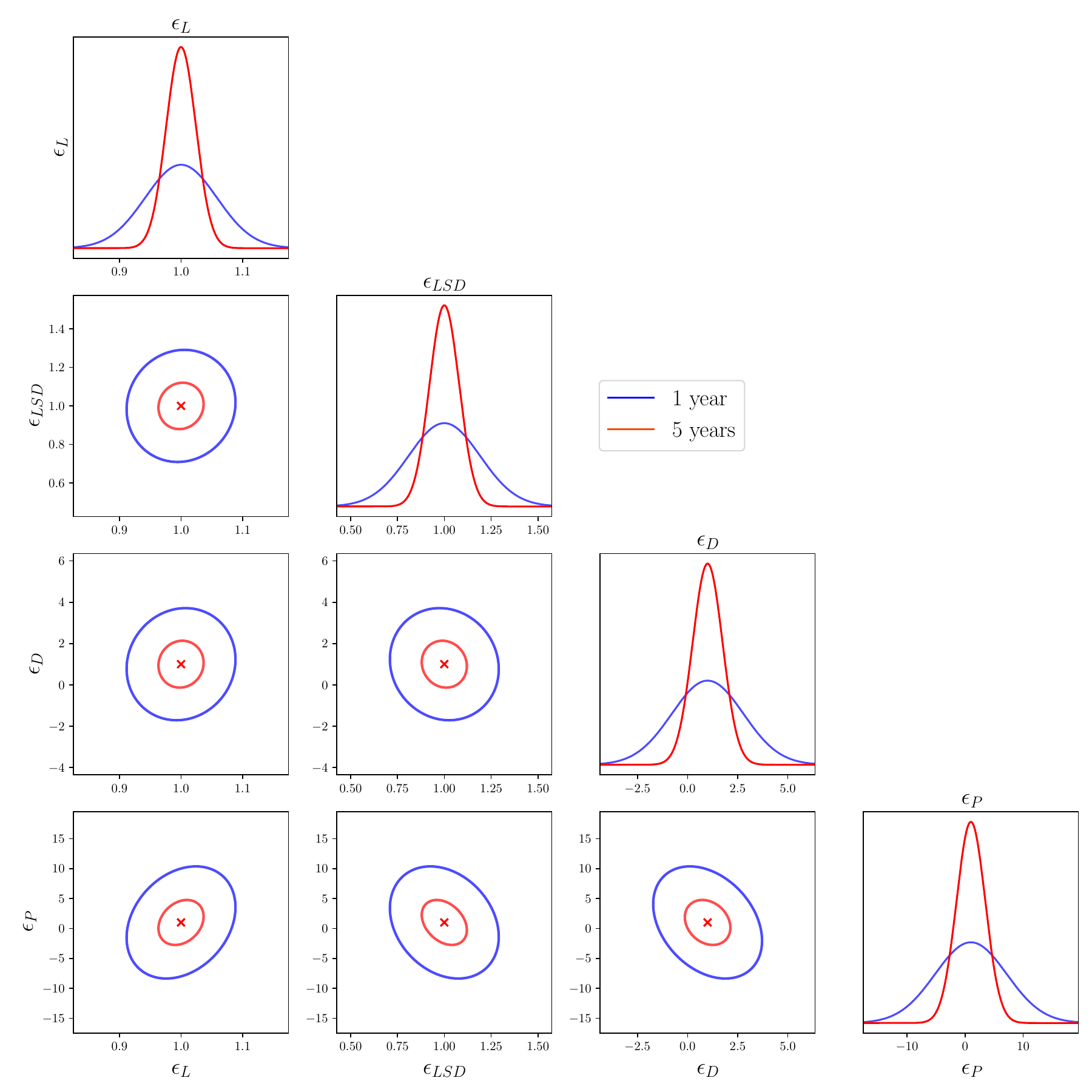}
    \caption{Predicted measurements of different relativistic effects to the observed number counts fluctuation in a cross-correlation of an ET-like GWs survey and an LSST-like galaxy survey. From top to bottom, we show forecasted measurement of lensing ($\epsilon_L$), LSD ($\epsilon_{LSD}$), Doppler ($\epsilon_D$) and GR corrections ($\epsilon_P$). As previously, we show both $1$ and $2\sigma$ contour, and the contour ellipses for $1$ and $5$ years of observations, respectively blue and red.}
    \label{fig:epsilons} 
\end{figure*}

We now proceed to forecast the detectability of different relativistic effects. We compute this by coupling each term in \autoref{eq:coefficients} to a dummy amplitude $\epsilon$ which we set to $1$, i.e., \be
\Delta=\Delta_g+\epsilon_{LSD}\Delta_{LSD}+\epsilon_{L}\Delta_{L}+\epsilon_{D}\Delta_{D}+\epsilon_{P}\Delta_{P}\,.
\ee
We forecast measurements of LSD, Lensing, Doppler and other GR corrections (mostly potential terms) - shown in \autoref{eq:coefficients} to \autoref{eq:AL} - by measuring if we can constrain the respective amplitudes $\epsilon_{LSD}$, $\epsilon_L$, $\epsilon_D$ and $\epsilon_P$; the latter is coupled to the remaining GR corrections in the term $g_{GW}(\Psi,\Phi,r)$ of  \autoref{eq:density_contrast}. We forecast these measurements for a cross-correlation of a third generation GW experiment such as ET with an LSST-like galaxy survey. We report the contour ellipses in \autoref{fig:epsilons}. 

The lensing bounds in \autoref{fig:epsilons} are particularly exciting. With simply $1$ year of observations, the confidence in the measured lensing amplitude is $\sim10\%$. However, with $5$ years of data the error in the detection shrinks to $\sim3\%$. This result indicates the clear possibility of measuring the lensing effect in luminosity distance space using GWs as tracers.

Interestingly, the radial velocity distortion $LSD$ effect is also measurable, albeit with worse precision. This is broadly due to the wide bins that can only be used in GWs. With a $5$-year observation period the error approaches $10\%$.

Regarding the other effects, i.e. Doppler and GR, the results are less optimistic, implying the measurements will be likely impossible. However, this is to be expected as they are all subdominant terms in the angular power spectrum, as shown by \citet{Us}.

Finally, we point out that whilst lensing, $LSD$ and Doppler show virtually no degeneracy between each other, the measurement of the remaining GR corrections seem correlated to all of them. However, this is simply an artefact of the axes limits, and the degeneracy angle is negligible.

\begin{figure*}
    \centering
    \includegraphics[width=\textwidth]{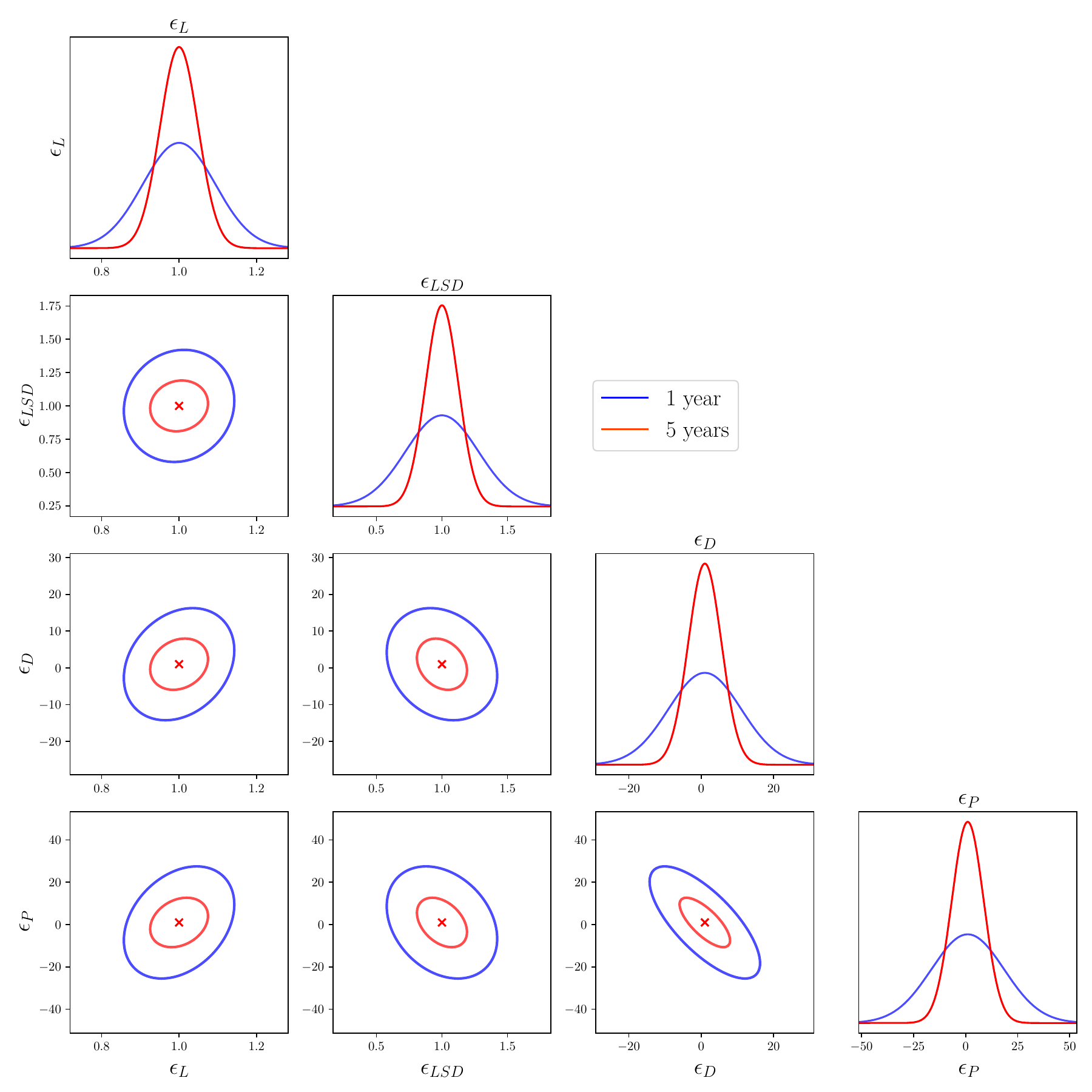}
    \caption{Predicted measurements of different relativistic effects to the observed number counts fluctuation in a cross-correlation of an ET-like GWs survey and a Megamapper-like galaxy survey. The description of the figure is the same as in \autoref{fig:epsilons} above.}
    \label{fig:epsi_mega}
\end{figure*}

We compute the same analysis by cross-correlating an ET-like experiment and a Megamapper-like survey, and report the results in \autoref{fig:epsi_mega}. We note immediately that the constraints on all parameters are worse than an ET-LSST-like. For $\epsilon_L$ and $\epsilon_{LSD}$, the errors worsen by a factor of $\sim 2$, whilst this increases dramatically for $\epsilon_D$ and $\epsilon_P$. This due to small overlap of the redshifts covered by the two different experiments. Whilst we assume a population of BBHs seen by ET between $0<z<3$, a Megamapper-like survey would target LBG galaxies at $z>2$. 
This lowers the constraining power of a multitracer analyses. In addition, the fact that Megamapper is at redshifts higher than ET, depresses the detectability of the lensing, as galaxies no longer trace the lenses in the line-of-sight of GWs.
Furthermore, the higher end of the redshift range covered by an ET-like detector sees higher shot noise due to the lower number of sources available. This is simply due to our assumption of a Madau-Dickinson rate in \autoref{eq:madau}, and a different population might result in separate noise. 

\subsection{Magnification and evolution bias}\label{sec:biases}
As per \autoref{sec:app_survey}, both $s$ and $b_e$ are functions of redshift, and are parametrised as cubic polynomials with coefficients $s_a,s_b,s_c,s_d$ and $b_{e,a},b_{e,b},b_{e,c},b_{e,d}$. 
However, in order not to reduce constraining power, as it would be spread across these coefficients, we opt to constrain the value of the biases at the centre of each redshift bin. Thus, we compute Fisher matrices including the parameters $\theta = \{B,\alpha,s^{GW}(z_k),b_e^{GW}(z_k),\theta_c\}$, where $z_k$ runs over the bins. 

\begin{figure}
    \centering
    \includegraphics[width=\linewidth]{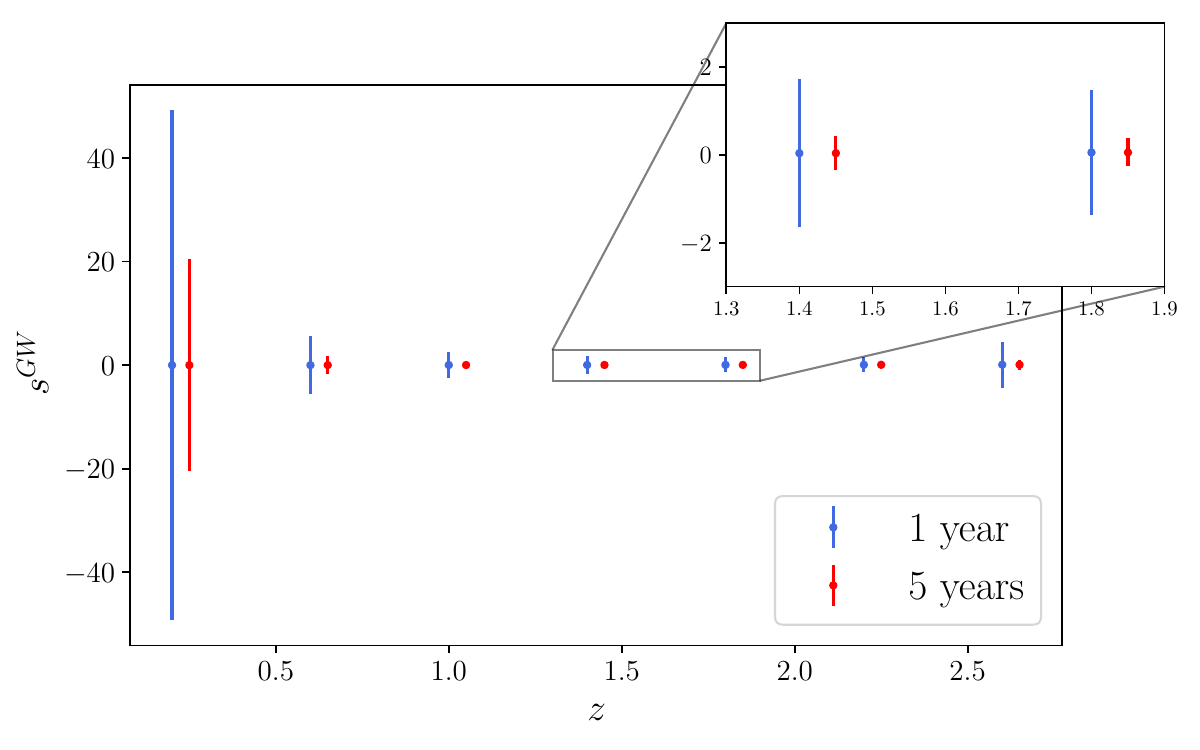}
    \includegraphics[width=\linewidth]{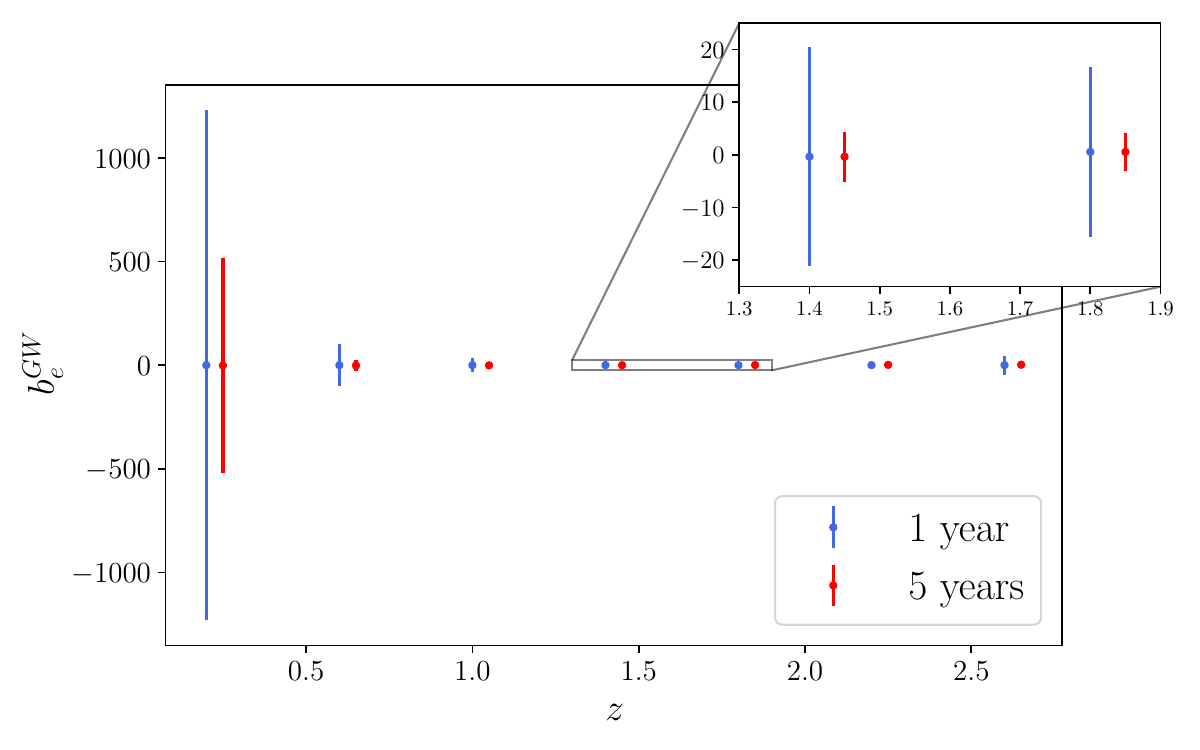}
    \caption{\textit{Left}: Forecasted measurement of the ET-like magnification bias $s^{GW}$. Blue errorbars indicate $1$ year of observation, red ones $5$ years. The latter are shifted horizontally for display purposes, however they are at the same redshift value as the blue points. \textit{Right:} Forecasted measurement of the ET-like evolution bias $b_e^{GW}$.}
    \label{fig:magbias}
\end{figure}

Considering the precision obtained in the previous section, we perform this analysis only with the combination of surveys which yielded the best results, i.e. an ET $\times$ LSST-like cross-correlation. 
We display our results in \autoref{fig:magbias}, showing the forecasted measure of an ET-like $s^{GW}$. Accurate redshift binning is crucial in this analysis. The more bins, the more measurement data points we obtain; however the shot noise increases as the number of sources in each bin drops. We find that bins of $\Delta z = 0.4$ provided a sufficient balance of the two effects considered, and is the binning provided in \autoref{fig:magbias}. Here we again show the measurement with $1$ year of observation (in blue) and with $5$ years (in red). We note that bins towards the centre of the observed redshift range, i.e. $z=1.4,z=1.8$, have the highest precision.

We performed a similar analysis for the evolution bias. However, the results yielded were much worse and suggest it may be impossible to measure $b_e$. 

\begin{figure}
    \centering
    \includegraphics[width=0.9\linewidth]{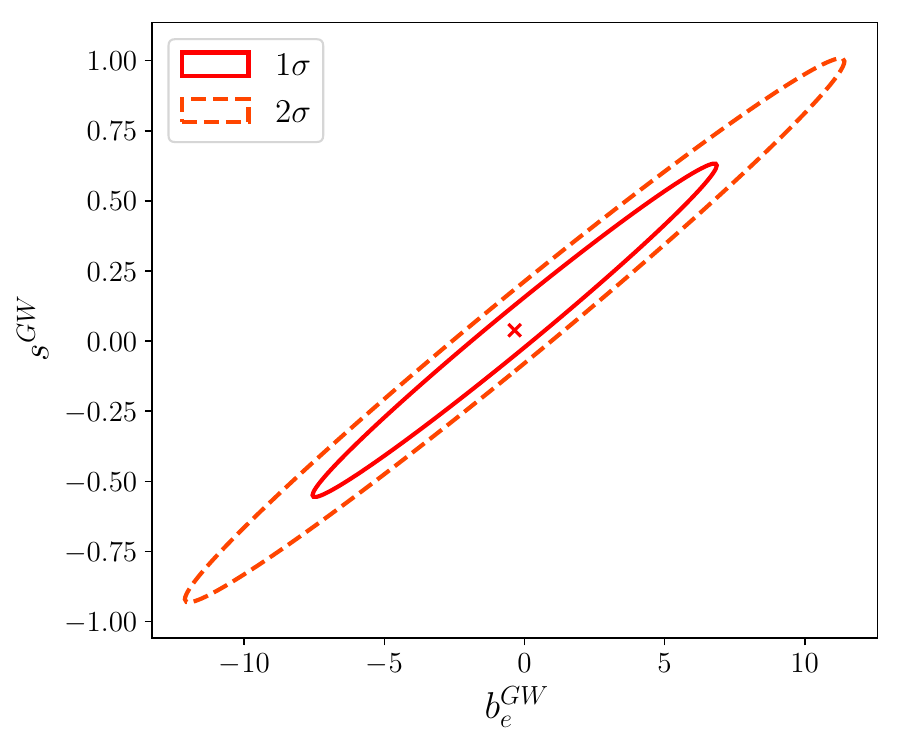}
    \caption{Forecasted measurement on the magnification bias $s^{GW}$ and evolution bias $b_e^{GW}$ at $z=1.4$. The degeneracy angle is $\sim 5\degree$ at this value of redshift.}
    \label{fig:ellipses}
\end{figure}

\begin{figure}
    \centering
    \includegraphics[width=\linewidth]{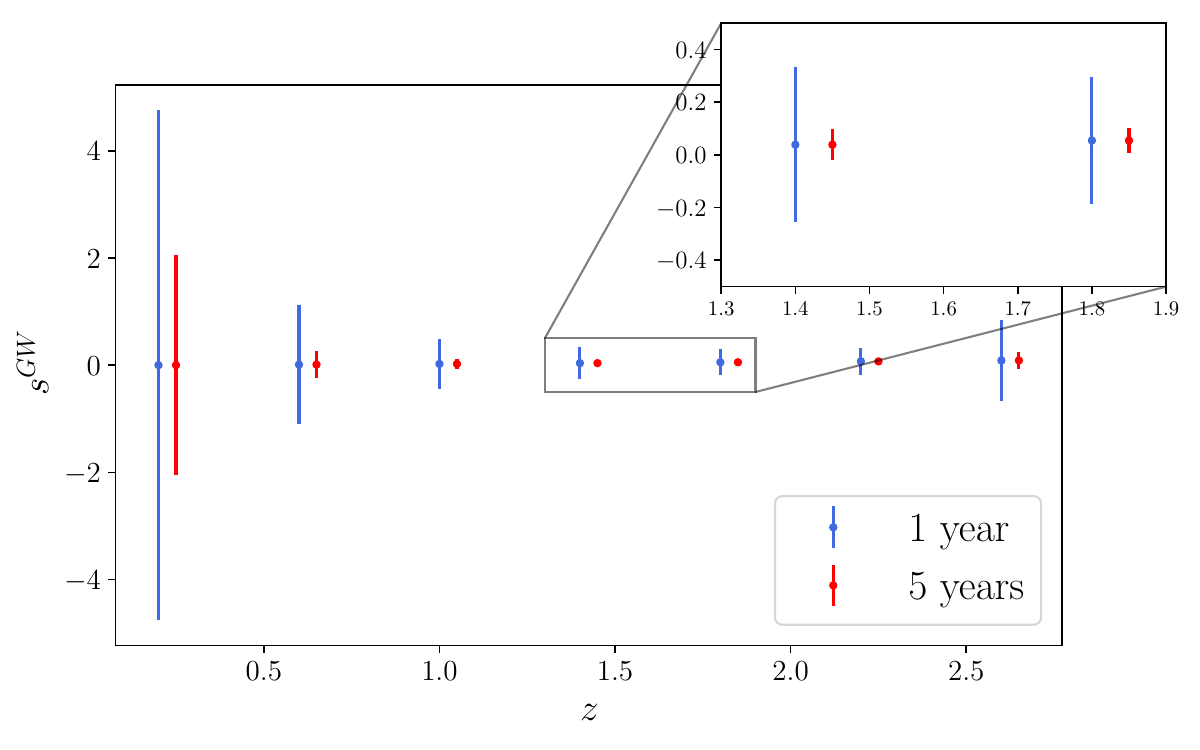}
    \includegraphics[width=\linewidth]{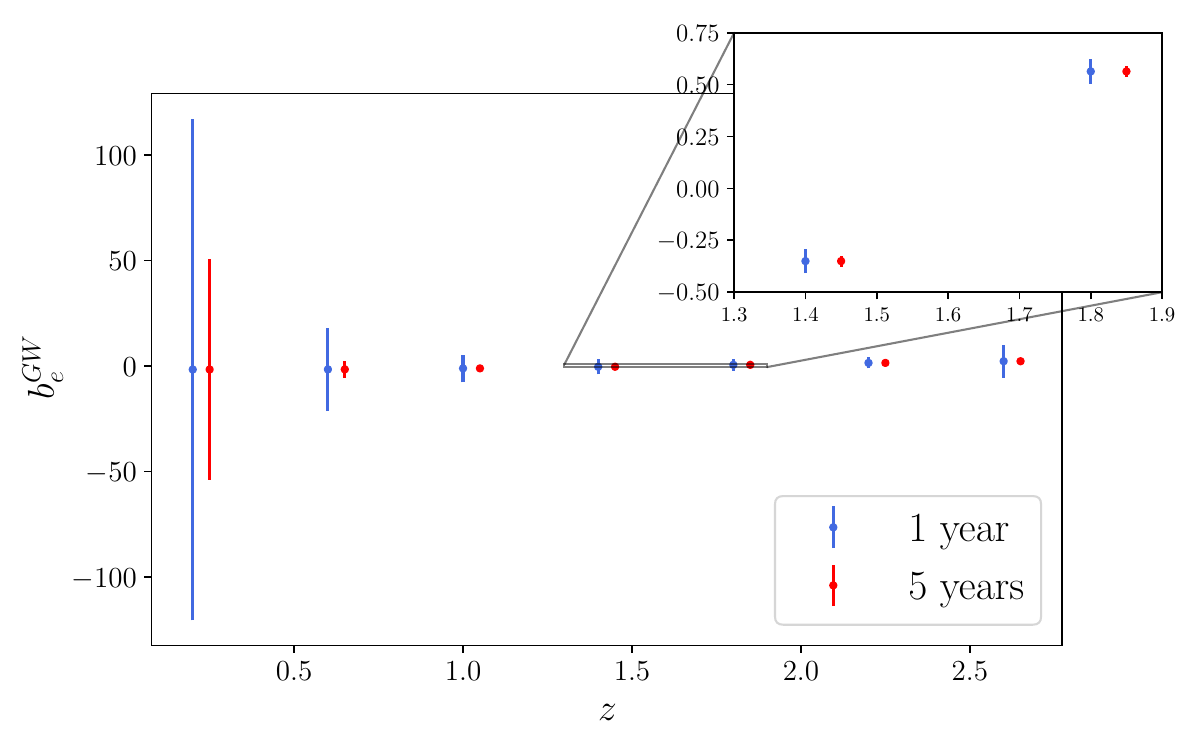}
    \caption{Forecasted measurements of the ET-like magnification (\textit{top}) bias by fixing the evolution, and viceversa (\textit{bottom}) As in \autoref{fig:magbias}, blue errorbars indicate $1$ year of observations, whilst red ones show $5$ years.}
    \label{fig:biases_notmarg}
\end{figure}

Both parameters appear in the lensing amplitude, which is one of the dominant contributions to the observed number counts fluctuation. The fact that the lensing term can be measured in cross-correlation, but the error on $s^{GW}$ and $b_e^{GW}$ is quite large, reveals the degeneracy between the two parameters. This can be noticed in the definition of $\beta$ (\autoref{eq:beta}), and explicitly in \autoref{fig:ellipses}, where we plot the forecast ellipse on the values of $s^{GW}$ and $b_e^{GW}$ at $z=1.4$. In essence, one can only measure a linear combination of the two parameters.
Therefore, in order to increase the precision on either term we can investigate what happens when we assume perfect knowledge of the other. Effectively, we fix one of the two parameters as opposed to marginalising over it. This method obviously yields a much higher precision, and in particular it produces a clear region $1<z<2$ where the measurement of either bias has better forecasted measured error bars, reaching percent level precision, as we show in the panels of \autoref{fig:biases_notmarg}. 

This particular method can be exploited to look for signatures of new physics, namely testing the equivalence principle. Euler's equation is normally used to construct the relativistic number counts, as it's done in \citet{Us}. However, if Euler's equation is violated, then the relativistic correction to the number counts can be recasted, as it was done in \citet{Bonvin_2018} in redshift space, notably in their equation 3.6. From such equation, one can note that deviations from the equivalence principle are then degenerate with a combination of both evolution and magnification bias. 

Ultimately, constraining a combination these parameters would help constrain the model of $n_{GW}$ in \autoref{eq:GW_numdens}, and in particular of the underlying merger rate $R_{GW}$. In fact, \citet{Zazzera_2024} found that the difference in the biases when using two chirp mass distributions $\phi(\mathcal{M})$ of BBH mergers from the GWTC-3 catalogue \citep{GWTC3_LVK,LIGOScientific}, i.e. either a \textit{Power Law + Peak} (as used in this paper) or \textit{Broken Power Law}, is negligible. Thus, if new detections constrain $\phi(\mathcal{M})$ further, then measuring these biases would effectively measure the underlying merger rate of BBHs. Further work in this direction would include varying the coefficients and amplitudes of the merger rate $R_{GW}$ in \autoref{eq:madau}, and assess the uncertainty in the measurement of the rate itself.

\section{Discussion and conclusions}\label{sec:conclusion}

In this paper, we presented forecasts on the precision of measuring clustering, magnification and evolution bias of gravitational waves from binary black holes. We established that a cross-correlation of current and/or near future surveys will not yield a measurement error lower than $\sim 25\%$, possibly due to the limited number of redshift bins accessible by an O$4$- or O$5$-like experiment. This puts a larger bound on the measured clustering bias as the density term of the angular power spectrum is probed very little.

In contrast, 3rd generation detectors will be able to access a vast redshift range, thus being able to probe the density term across different redshifts. Further, shot noise is greatly reduced with ET-like detectors given the large (predicted) increase in detections from O$5$. 

In fact, this is made evident in \autoref{fig:timeline}, showing how precision in measuring the amplitude of the clustering bias of GWs shrinks from $\sim50\%$ with an O$4\times$DESI-like cross-correlation, to $2.5\%$ with an ET$\times$LSST-like one. Further, reducing the shot noise with longer observations, thus higher number of detections, clearly yields increased precision.

This type of cross-correlation will then be able to measure the relation between BBHs which live in DM halos and the DM distribution itself, i.e. the clustering bias. However, GWs from isolated binaries, or from binary compact objects which are not following the underlying DM distribution such as primordial black holes (PBHs) would not be following this trend. Instead, would likely result in increased noise in \autoref{fig:clus_bias_ratio}. 

An ET$\times$LSST-like cross-correlation will also allow us to measure the lensing effect in the number counts of GWs due to galaxies with percent level precision, opening up to exciting opportunities to test gravity in large scale structure studies with GWs. Different modified gravity theories would result in an alternative expression for the number counts (as investigated in e.g. \citet{Balaudo_2024}), especially affecting terms such as the lensing term. Measuring the latter would then help constrain different modifications of gravity. 

Despite the precision obtained on the measurement of the lensing effect in the density contrast, 
we found a degeneracy in the measurements of the magnification and evolution biases, leading to large uncertainties on each parameter when marginalising over the other one. Assuming knowledge on either, i.e. keeping it as fixed, reduces the uncertainty drastically, yielding good measurements of both biases. In particular, we highlight an optimal region at $1<z<2.5$ where the error bars on both parameters are smallest. This is likely a combination of two contrasting effects. On one side, the higher the distance to us, the larger the lensing effect (which is the leading term including the biases) will be; on the other hand, shot noise increases drastically after $z>2$ in our analysis, given our simple assumption of a standard Madau-Dickinson rate describing the underlying merger rate of BBHs. This peaks at intermediate redshifts ($z\sim1.5$), and falls off at higher $z$, resulting in fewer sources as we approach $z=3$, thus higher shot noise. A different merger rate will therefore impact the measurement precision, although it would require modifying both $s^{GW}$ and $b_e^{GW}$, both of which depend on the BBH merger rate. Changing the chirp mass distribution would alter this as well. However, current LVK data suggest two most likely forms of primary mass distribution, \textit{Power-Law + Peak} (used in this paper) and \textit{Broken Power-Law}. The two are very similar, and \citet{Zazzera_2024} showed the two results in almost identical expressions of the biases, thus having a much less significant impact. Further, as discussed above, a different population of binary compact objects producing GWs will result in different bias parameters. These populations of BBHs would then have to be treated differently, just as the different galaxy types targeted by the galaxy surveys examined here.

To conclude, we remark that cross-correlations with GWs in $D_L-$space and galaxies in $z-$space will yield precise measurements of the clustering bias of GWs when third generation detectors such as ET will come online. The lensing effect will also be measurable to a $\sim3\%$ error, and the amplitude of luminosity space distortions to $\sim10\%$.

\section*{Acknowledgements}
We are pleased to thank Roy Maartens for useful discussions and Samantha Rossiter for providing helpful parameterisations of galaxy number densities. 
S.Z. acknowledges support from the Perren Fund, University of London. JF acknowledges support of Funda\c{c}\~{a}o para a Ci\^{e}ncia e a Tecnologia through the Investigador FCT Contract No. 2020.02633.CEECIND/CP1631/CT0002, the FCT project PTDC/FIS-AST/0054/2021, and the research grants UIDB/04434/2020 and UIDP/04434/2020. T.B. is supported by ERC Starting Grant \textit{SHADE} (grant no.~StG 949572) and a Royal Society University Research Fellowship (grant no.~URF$\backslash$R$\backslash$231006).

\appendix

\section{Surveys' specifications}\label{sec:specs}

\begin{itemize}
   \item \textit{DESI BGS-like}:
    The BGS number density is expected to closely follow the GAMA survey (\cite{DESIp1,Ruiz_Macias_2021}) and resulting fitting functions are given by \cite{Maartens_2021}:
    \bea
        n_{BGS}(z) &=& 0.023z^{-0.471}\exp{(-5.17z)}-0.002 \, ,\\
        b(z) &=& \frac{1.7}{f(z)} \, ,
    \eea
    where $f(z)$ is the growth rate, and $n_{BGS}$ is in $(h/\text{Mpc})^{3}$ units.\vspace{1mm}
    We model the required parameters using fitting functions from \cite{Maartens_2021}. The redshift range covered is $[0.9,1.6]$ \citep{Euclid_2019} with parametrisations
    \bea
        n_{H\alpha}(z) &=& 3.63z^{-0.91}\exp{(0.402z)}-4.14)\times10^{-3} \, ,\\
        b(z) &=& 0.7(1+z) \, .
    \eea
    Note that $n_{H\alpha}$ is in $(h/\text{Mpc})^{3}$ units.
    \item \textit{LSST-like:}
    For a photometric LSST-like survey, we take the luminosity function $\Phi$ of the target galaxies from \cite{LSST_book,LSST_2012, LSST_2021}:
    \bea
        n_{g}(z) &=& \int_{-\infty}^{M_{cut}}{\rm{d}}M\ \mathcal{F}(z,M) \, ,\\
        \mathcal{F}(z,M) &=& \left(\frac{\log10}{2.5}\right)\phi^*(z)10^{[0.4(M^*-M)]^{\alpha+1}}\times\nn\\
        &&\exp\left[10^{0.4(M^*-M)}\right] \,\\
        \phi^*(z) &=& (2.59-0.136z-0.081z^2)10^{-3} \,\\
        b &=& 1 + 0.84z \, ,
    \eea
    where for the full LSST sample we have $\alpha=-1.33$ and $M*=-21.49-1.25\log(1+z)$. 
    \item \textit{Megamapper-like}:
    Finally, for Lyman-break Galaxies (LBG) in the redshift range $2\lesssim z\lesssim 5$, we follow \citet{Astro2020,Sailer_2021,Megamapper}:
    \bea
        n_g(z) &=& \int_{-\infty}^{M_c}{\rm{d} M} \,\mathcal{F}(z,M) \, ,\\
        \mathcal{F}(z,M_{UV}) &=& \left(\frac{\log 10}{2.5}\right)\phi^*10^{-0.4(1+\alpha)(M_{UV}-M^*_{UV})}\times \nn\\
        &\times& \exp{\left(-10^{-0.4(M_{UV}-M^*_{UV}}\right)} \, , \\
        b(z) &=& 0.6(1+z)+0.11(1+z)^2 \, .\
    \eea
    We note that the absolute magnitude cut is defined as $M_c = 24.5 - 5\log_{10}\left(D_L(z)/{10\text{pc}}\right)+2.5\log_{10}(1+z)$. We use best fit values of $\alpha$, $\phi*$ and $M_{UV}$ from table 3 of \citet{Wilson_2019} for each redshift bin.
    \item \textit{BBHs}: in \autoref{tab:GWsmagbiases} we summarise the amplitudes of a polynomial fit of the form:$a+bz+cz^2+dz^3$, to the magnification and evolution bias of different GWs experiments.
\end{itemize}

\begin{table}
    \centering
    {\setlength{\extrarowheight}{3pt}
    \begin{tabular}{|c|c|c|c|c|c|c|}
    \hline
         & \multicolumn{2}{c|}{O4-like} & \multicolumn{2}{c|}{O5-like} & \multicolumn{2}{c|}{ET-like} \\
    \hline
         & $s^{GW}$ & $b_e^{GW}$ & $s^{GW}$ & $b_e^{GW}$ & $s^{GW}$ & $b_e^{GW}$ \\
    \hline
         $a$ & -0.209 & -0.805 & -0.118 & -1.19 & \num{-3.77e-3} & \num{-1.46} \\
    \hline
         $b$ & 3.82 & -16.0 & 1.43 & -5.29 & \num{1.97e-2} & \num{-1.29} \\
    \hline
         $c$ & -0.138 & 58.2 & -2.42 & 10.4 & \num{9.43e-3} & 2.00 \\
    \hline
         $d$ & 21.9 & -92.3 & 1.62 & -6.53 & \num{-1.34e-3} & $-0.366$ \\
    \hline
    \end{tabular}}
    \caption{Coefficients of the cubic polynomial fits to the magnification and evolution biases for the different GW experiments. Each bias follows a cubic spline in redshift of form $a+bz+cz^2+dz^3$. Taken from \citet{Zazzera_2024}.}
    \label{tab:GWsmagbiases}
\end{table}

\bibliographystyle{mnras} 
\bibliography{forecastbib} 

\begin{thebibliography}{}
\makeatletter
\relax
\def\mn@urlcharsother{\let\do\@makeother \do\$\do\&\do\#\do\^\do\_\do\%\do\~}
\def\mn@doi{\begingroup\mn@urlcharsother \@ifnextchar [ {\mn@doi@} {\mn@doi@[]}}
\def\mn@doi@[#1]#2{\def\@tempa{#1}\ifx\@tempa\@empty \href {http://dx.doi.org/#2} {doi:#2}\else \href {http://dx.doi.org/#2} {#1}\fi \endgroup}
\def\mn@eprint#1#2{\mn@eprint@#1:#2::\@nil}
\def\mn@eprint@arXiv#1{\href {http://arxiv.org/abs/#1} {{\tt arXiv:#1}}}
\def\mn@eprint@dblp#1{\href {http://dblp.uni-trier.de/rec/bibtex/#1.xml} {dblp:#1}}
\def\mn@eprint@#1:#2:#3:#4\@nil{\def\@tempa {#1}\def\@tempb {#2}\def\@tempc {#3}\ifx \@tempc \@empty \let \@tempc \@tempb \let \@tempb \@tempa \fi \ifx \@tempb \@empty \def\@tempb {arXiv}\fi \@ifundefined {mn@eprint@\@tempb}{\@tempb:\@tempc}{\expandafter \expandafter \csname mn@eprint@\@tempb\endcsname \expandafter{\@tempc}}}

\bibitem[\protect\citeauthoryear{Aasi et~al.,}{Aasi et~al.}{2015}]{AdvLIGO}
Aasi J.,  et~al., 2015, \mn@doi [Classical and Quantum Gravity] {10.1088/0264-9381/32/7/074001}, 32, 074001

\bibitem[\protect\citeauthoryear{Abbott et~al.}{Abbott et~al.}{2016a}]{KAGRA:2013rdx}
Abbott B.~P.,  et~al., 2016a, \mn@doi [Living Rev. Rel.] {10.1007/s41114-020-00026-9}, 19, 1

\bibitem[\protect\citeauthoryear{Abbott et~al.,}{Abbott et~al.}{2016b}]{GWfirst}
Abbott B.~P.,  et~al., 2016b, \mn@doi [Phys. Rev. Lett.] {10.1103/PhysRevLett.116.061102}, 116, 061102

\bibitem[\protect\citeauthoryear{Abbott et~al.,}{Abbott et~al.}{2021}]{GWTC2}
Abbott R.,  et~al., 2021, \mn@doi [The Astrophysical Journal Letters] {10.3847/2041-8213/abe949}, 913, L7

\bibitem[\protect\citeauthoryear{Abbott et~al.,}{Abbott et~al.}{2023}]{Abbott_2023}
Abbott R.,  et~al., 2023, \mn@doi [The Astrophysical Journal] {10.3847/1538-4357/ac74bb}, 949, 76

\bibitem[\protect\citeauthoryear{Abramo, Dinarte~Ferri, Tashiro  \& Loureiro}{Abramo et~al.}{2022}]{Abramo_2022}
Abramo L.~R.,  Dinarte~Ferri J.~V.,  Tashiro I.~L.,   Loureiro A.,  2022, \mn@doi [Journal of Cosmology and Astroparticle Physics] {10.1088/1475-7516/2022/08/073}, 2022, 073

\bibitem[\protect\citeauthoryear{Acernese et~al.,}{Acernese et~al.}{2014}]{AdvVirgo}
Acernese F.,  et~al., 2014, \mn@doi [Classical and Quantum Gravity] {10.1088/0264-9381/32/2/024001}, 32, 024001

\bibitem[\protect\citeauthoryear{Afroz \& Mukherjee}{Afroz \& Mukherjee}{2024}]{Afroz:2024joi}
Afroz S.,  Mukherjee S.,  2024, \mn@doi [Mon. Not. Roy. Astron. Soc.] {10.1093/mnras/stae2139}, 534, 1283

\bibitem[\protect\citeauthoryear{Akutsu et~al.,}{Akutsu et~al.}{2020}]{advKagra}
Akutsu T.,  et~al., 2020, Overview of KAGRA: Detector design and construction history (\mn@eprint {arXiv} {2005.05574}), \url {https://arxiv.org/abs/2005.05574}

\bibitem[\protect\citeauthoryear{Balaudo, Garoffolo, Martinelli, Mukherjee  \& Silvestri}{Balaudo et~al.}{2023}]{Balaudo:2022znx}
Balaudo A.,  Garoffolo A.,  Martinelli M.,  Mukherjee S.,   Silvestri A.,  2023, \mn@doi [JCAP] {10.1088/1475-7516/2023/06/050}, 06, 050

\bibitem[\protect\citeauthoryear{Balaudo, Pantiri  \& Silvestri}{Balaudo et~al.}{2024}]{Balaudo_2024}
Balaudo A.,  Pantiri M.,   Silvestri A.,  2024, Number count of Gravitational Waves and Supernovae in Luminosity Distance space for LCDM and Scalar-Tensor theories (\mn@eprint {arXiv} {2311.17904}), \url {https://arxiv.org/abs/2311.17904}

\bibitem[\protect\citeauthoryear{Blanchard et~al.,}{Blanchard et~al.}{2020}]{Euclid_2019}
Blanchard A.,  et~al., 2020, \mn@doi [Astronomy &amp; Astrophysics] {10.1051/0004-6361/202038071}, 642, A191

\bibitem[\protect\citeauthoryear{Bonvin}{Bonvin}{2008}]{Bonvin_2008}
Bonvin C.,  2008, \mn@doi [Physical Review D] {10.1103/physrevd.78.123530}, 78

\bibitem[\protect\citeauthoryear{Bonvin \& Durrer}{Bonvin \& Durrer}{2011}]{Bonvin_2011}
Bonvin C.,  Durrer R.,  2011, \mn@doi [Physical Review D] {10.1103/physrevd.84.063505}, 84

\bibitem[\protect\citeauthoryear{Bonvin \& Fleury}{Bonvin \& Fleury}{2018}]{Bonvin_2018}
Bonvin C.,  Fleury P.,  2018, \mn@doi [Journal of Cosmology and Astroparticle Physics] {10.1088/1475-7516/2018/05/061}, 2018, 061–061

\bibitem[\protect\citeauthoryear{Calore, Cuoco, Regimbau, Sachdev  \& Serpico}{Calore et~al.}{2020}]{Calore_2020}
Calore F.,  Cuoco A.,  Regimbau T.,  Sachdev S.,   Serpico P.~D.,  2020, \mn@doi [Physical Review Research] {10.1103/physrevresearch.2.023314}, 2

\bibitem[\protect\citeauthoryear{Castorina \& Dio}{Castorina \& Dio}{2022}]{Castorina_2022}
Castorina E.,  Dio E.~D.,  2022, \mn@doi [Journal of Cosmology and Astroparticle Physics] {10.1088/1475-7516/2022/01/061}, 2022, 061

\bibitem[\protect\citeauthoryear{Challinor \& Lewis}{Challinor \& Lewis}{2011}]{Challinor_2011}
Challinor A.,  Lewis A.,  2011, \mn@doi [Physical Review D] {10.1103/physrevd.84.043516}, 84

\bibitem[\protect\citeauthoryear{Chen}{Chen}{2008}]{Chen_2008}
Chen J.,  2008, \mn@doi [A\&A] {10.1051/0004-6361:20079184}, 494, 867

\bibitem[\protect\citeauthoryear{Cigarrán Díaz \& Mukherjee}{Cigarrán Díaz \& Mukherjee}{2022}]{Cigarr_n_D_az_2022}
Cigarrán Díaz C.,  Mukherjee S.,  2022, \mn@doi [Monthly Notices of the Royal Astronomical Society] {10.1093/mnras/stac208}, 511, 2782–2795

\bibitem[\protect\citeauthoryear{Evans et~al.,}{Evans et~al.}{2021}]{Evans_2021}
Evans M.,  et~al., 2021, A Horizon Study for Cosmic Explorer: Science, Observatories, and Community (\mn@eprint {arXiv} {2109.09882})

\bibitem[\protect\citeauthoryear{Finn}{Finn}{1996}]{Finn_1996}
Finn L.~S.,  1996, \mn@doi [Physical Review D] {10.1103/physrevd.53.2878}, 53, 2878

\bibitem[\protect\citeauthoryear{Fonseca, Zazzera, Baker  \& Clarkson}{Fonseca et~al.}{2023}]{Us}
Fonseca J.,  Zazzera S.,  Baker T.,   Clarkson C.,  2023, The observed number counts in luminosity distance space (\mn@eprint {arXiv} {2304.14253})

\bibitem[\protect\citeauthoryear{Hui, Gazta{\~{n} }aga  \& LoVerde}{Hui et~al.}{2008}]{Hui_2008}
Hui L.,  Gazta{\~{n} }aga E.,   LoVerde M.,  2008, \mn@doi [Physical Review D] {10.1103/physrevd.77.063526}, 77

\bibitem[\protect\citeauthoryear{Jeong, Schmidt  \& Hirata}{Jeong et~al.}{2012}]{Jeong_2012}
Jeong D.,  Schmidt F.,   Hirata C.~M.,  2012, \mn@doi [Physical Review D] {10.1103/physrevd.85.023504}, 85

\bibitem[\protect\citeauthoryear{{LIGO-VIRGO-KAGRA Collaboration} et~al.,}{{LIGO-VIRGO-KAGRA Collaboration} et~al.}{2022}]{GWTC3_LVK}
{LIGO-VIRGO-KAGRA Collaboration} R.~A.,  et~al., 2022, The population of merging compact binaries inferred using gravitational waves through GWTC-3 (\mn@eprint {arXiv} {2111.03634})

\bibitem[\protect\citeauthoryear{{LIGO-VIRGO-KAGRA collaboration} et~al.}{{LIGO-VIRGO-KAGRA collaboration} et~al.}{2021}]{LIGOScientific}
{LIGO-VIRGO-KAGRA collaboration} Abbott R.,  et~al., 2021, GWTC-3: Compact Binary Coalescences Observed by LIGO and Virgo During the Second Part of the Third Observing Run (\mn@eprint {arXiv} {2111.03606})

\bibitem[\protect\citeauthoryear{{LSST Dark Energy Science Collaboration}}{{LSST Dark Energy Science Collaboration}}{2012}]{LSST_2012}
{LSST Dark Energy Science Collaboration} 2012, Large Synoptic Survey Telescope: Dark Energy Science Collaboration (\mn@eprint {arXiv} {1211.0310}), \url {https://arxiv.org/abs/1211.0310}

\bibitem[\protect\citeauthoryear{{LSST Science Collaboration} et~al.,}{{LSST Science Collaboration} et~al.}{2009}]{LSST_book}
{LSST Science Collaboration} et~al., 2009, LSST Science Book, Version 2.0 (\mn@eprint {arXiv} {0912.0201})

\bibitem[\protect\citeauthoryear{{LSST Science Collaboration} et~al.,}{{LSST Science Collaboration} et~al.}{2021}]{LSST_2021}
{LSST Science Collaboration} B.~A.,  et~al., 2021, \mn@doi [The Astrophysical Journal Supplement Series] {10.3847/1538-4365/abd62c}, 253, 31

\bibitem[\protect\citeauthoryear{Laureijs et~al.,}{Laureijs et~al.}{2011}]{EuclidOG}
Laureijs R.,  et~al., 2011, Euclid Definition Study Report (\mn@eprint {arXiv} {1110.3193}), \url {https://arxiv.org/abs/1110.3193}

\bibitem[\protect\citeauthoryear{Levi et~al.,}{Levi et~al.}{2013}]{DESIwp}
Levi M.,  et~al., 2013, The DESI Experiment, a whitepaper for Snowmass 2013 (\mn@eprint {arXiv} {1308.0847}), \url {https://arxiv.org/abs/1308.0847}

\bibitem[\protect\citeauthoryear{Libanore et~al.,}{Libanore et~al.}{2021}]{Libanore_2021}
Libanore S.,  et~al., 2021, \mn@doi [Journal of Cosmology and Astroparticle Physics] {10.1088/1475-7516/2021/02/035}, 2021, 035

\bibitem[\protect\citeauthoryear{Libanore, Artale, Karagiannis, Liguori, Bartolo, Bouffanais, Mapelli  \& Matarrese}{Libanore et~al.}{2022}]{Libanore_2022}
Libanore S.,  Artale M.,  Karagiannis D.,  Liguori M.,  Bartolo N.,  Bouffanais Y.,  Mapelli M.,   Matarrese S.,  2022, \mn@doi [Journal of Cosmology and Astroparticle Physics] {10.1088/1475-7516/2022/02/003}, 2022, 003

\bibitem[\protect\citeauthoryear{Ma, Hu  \& Huterer}{Ma et~al.}{2006}]{Ma_2006}
Ma Z.,  Hu W.,   Huterer D.,  2006, \mn@doi [The Astrophysical Journal] {10.1086/497068}, 636, 21–29

\bibitem[\protect\citeauthoryear{Maartens, Fonseca, Camera, Jolicoeur, Viljoen  \& Clarkson}{Maartens et~al.}{2021}]{Maartens_2021}
Maartens R.,  Fonseca J.,  Camera S.,  Jolicoeur S.,  Viljoen J.-A.,   Clarkson C.,  2021, \mn@doi [Journal of Cosmology and Astroparticle Physics] {10.1088/1475-7516/2021/12/009}, 2021, 009

\bibitem[\protect\citeauthoryear{Madau \& Dickinson}{Madau \& Dickinson}{2014}]{MadauDickinson}
Madau P.,  Dickinson M.,  2014, \mn@doi [Annual Review of Astronomy and Astrophysics] {10.1146/annurev-astro-081811-125615}, 52, 415

\bibitem[\protect\citeauthoryear{Maggiore et~al.,}{Maggiore et~al.}{2020}]{Maggiore_2020}
Maggiore M.,  et~al., 2020, \mn@doi [Journal of Cosmology and Astroparticle Physics] {10.1088/1475-7516/2020/03/050}, 2020, 050

\bibitem[\protect\citeauthoryear{Martinez, Merchan, Valotto  \& Lambas}{Martinez et~al.}{1999}]{Martinez_1999}
Martinez H.~J.,  Merchan M.~E.,  Valotto C.~A.,   Lambas D.~G.,  1999, \mn@doi [The Astrophysical Journal] {10.1086/306973}, 514, 558

\bibitem[\protect\citeauthoryear{Mastrogiovanni, Karathanasis, Gair, Ashton, Rinaldi, Huang  \& Dálya}{Mastrogiovanni et~al.}{2024}]{Mastrogiovanni_24}
Mastrogiovanni S.,  Karathanasis C.,  Gair J.,  Ashton G.,  Rinaldi S.,  Huang H.-Y.,   Dálya G.,  2024, \mn@doi [Annalen der Physik] {https://doi.org/10.1002/andp.202200180}, 536, 2200180

\bibitem[\protect\citeauthoryear{Mpetha, Congedo  \& Taylor}{Mpetha et~al.}{2023}]{Mpetha_2022}
Mpetha C.~T.,  Congedo G.,   Taylor A.,  2023, \mn@doi [Physical Review D] {10.1103/physrevd.107.103518}, 107

\bibitem[\protect\citeauthoryear{Mukherjee \& Wandelt}{Mukherjee \& Wandelt}{2018}]{mukherjee2018}
Mukherjee S.,  Wandelt B.~D.,  2018, Beyond the classical distance-redshift test: cross-correlating redshift-free standard candles and sirens with redshift surveys (\mn@eprint {arXiv} {1808.06615}), \url {https://arxiv.org/abs/1808.06615}

\bibitem[\protect\citeauthoryear{Mukherjee, Wandelt  \& Silk}{Mukherjee et~al.}{2020a}]{Mukherjee:2019wfw}
Mukherjee S.,  Wandelt B.~D.,   Silk J.,  2020a, \mn@doi [Phys. Rev. D] {10.1103/PhysRevD.101.103509}, 101, 103509

\bibitem[\protect\citeauthoryear{Mukherjee, Wandelt  \& Silk}{Mukherjee et~al.}{2020b}]{Mukherjee:2019wcg}
Mukherjee S.,  Wandelt B.~D.,   Silk J.,  2020b, \mn@doi [Mon. Not. Roy. Astron. Soc.] {10.1093/mnras/staa827}, 494, 1956

\bibitem[\protect\citeauthoryear{Mukherjee, Wandelt, Nissanke  \& Silvestri}{Mukherjee et~al.}{2021}]{Mukherjee_2021}
Mukherjee S.,  Wandelt B.~D.,  Nissanke S.~M.,   Silvestri A.,  2021, \mn@doi [Physical Review D] {10.1103/physrevd.103.043520}, 103

\bibitem[\protect\citeauthoryear{Mukherjee, Krolewski, Wandelt  \& Silk}{Mukherjee et~al.}{2024}]{Mukherjee:2022afz}
Mukherjee S.,  Krolewski A.,  Wandelt B.~D.,   Silk J.,  2024, \mn@doi [Astrophys. J.] {10.3847/1538-4357/ad7d90}, 975, 189

\bibitem[\protect\citeauthoryear{Oguri}{Oguri}{2018}]{Oguri_2018}
Oguri M.,  2018, \mn@doi [Monthly Notices of the Royal Astronomical Society] {10.1093/mnras/sty2145}, 480, 3842

\bibitem[\protect\citeauthoryear{Padmanabhan et~al.,}{Padmanabhan et~al.}{2007}]{Padmanabhan_2007}
Padmanabhan N.,  et~al., 2007, \mn@doi [Monthly Notices of the Royal Astronomical Society] {10.1111/j.1365-2966.2007.11593.x}, 378, 852

\bibitem[\protect\citeauthoryear{{Planck Collaboration} Aghanim et~al.,}{{Planck Collaboration} et~al.}{2020}]{Planck_2018}
{Planck Collaboration} Aghanim N.,  et~al., 2020, \mn@doi [Astronomy &amp; Astrophysics] {10.1051/0004-6361/201833910}, 641, A6

\bibitem[\protect\citeauthoryear{Punturo et~al.,}{Punturo et~al.}{2010}]{Punturo_2010}
Punturo M.,  et~al., 2010, \mn@doi [Classical and Quantum Gravity] {10.1088/0264-9381/27/8/084007}, 27, 084007

\bibitem[\protect\citeauthoryear{Reitze et~al.,}{Reitze et~al.}{2019}]{Reitze_2019}
Reitze D.,  et~al., 2019, Cosmic Explorer: The U.S. Contribution to Gravitational-Wave Astronomy beyond LIGO (\mn@eprint {arXiv} {1907.04833})

\bibitem[\protect\citeauthoryear{Ruiz-Macias et~al.,}{Ruiz-Macias et~al.}{2021}]{Ruiz_Macias_2021}
Ruiz-Macias O.,  et~al., 2021, \mn@doi [Monthly Notices of the Royal Astronomical Society] {10.1093/mnras/stab292}, 502, 4328–4349

\bibitem[\protect\citeauthoryear{Sailer, Castorina, Ferraro  \& White}{Sailer et~al.}{2021}]{Sailer_2021}
Sailer N.,  Castorina E.,  Ferraro S.,   White M.,  2021, \mn@doi [Journal of Cosmology and Astroparticle Physics] {10.1088/1475-7516/2021/12/049}, 2021, 049

\bibitem[\protect\citeauthoryear{Sanchez et~al.,}{Sanchez et~al.}{2022}]{Sanchez_2022}
Sanchez B.~O.,  et~al., 2022, \mn@doi [The Astrophysical Journal] {10.3847/1538-4357/ac7a37}, 934, 96

\bibitem[\protect\citeauthoryear{Sathyaprakash, Schutz  \& Broeck}{Sathyaprakash et~al.}{2010}]{Sathyaprakash_2010}
Sathyaprakash B.~S.,  Schutz B.~F.,   Broeck C. V.~D.,  2010, \mn@doi [Classical and Quantum Gravity] {10.1088/0264-9381/27/21/215006}, 27, 215006

\bibitem[\protect\citeauthoryear{Sathyaprakash et~al.,}{Sathyaprakash et~al.}{2012}]{Sathyaprakash_2012}
Sathyaprakash B.,  et~al., 2012, \mn@doi [Classical and Quantum Gravity] {10.1088/0264-9381/29/12/124013}, 29, 124013

\bibitem[\protect\citeauthoryear{Scelfo, Bellomo, Raccanelli, Matarrese  \& Verde}{Scelfo et~al.}{2018}]{Scelfo_2018}
Scelfo G.,  Bellomo N.,  Raccanelli A.,  Matarrese S.,   Verde L.,  2018, \mn@doi [Journal of Cosmology and Astroparticle Physics] {10.1088/1475-7516/2018/09/039}, 2018, 039

\bibitem[\protect\citeauthoryear{Scelfo, Boco, Lapi  \& Viel}{Scelfo et~al.}{2020}]{Scelfo_2020}
Scelfo G.,  Boco L.,  Lapi A.,   Viel M.,  2020, \mn@doi [Journal of Cosmology and Astroparticle Physics] {10.1088/1475-7516/2020/10/045}, 2020, 045

\bibitem[\protect\citeauthoryear{Scelfo, Spinelli, Raccanelli, Boco, Lapi  \& Viel}{Scelfo et~al.}{2022}]{Scelfo_2022}
Scelfo G.,  Spinelli M.,  Raccanelli A.,  Boco L.,  Lapi A.,   Viel M.,  2022, \mn@doi [Journal of Cosmology and Astroparticle Physics] {10.1088/1475-7516/2022/01/004}, 2022, 004

\bibitem[\protect\citeauthoryear{Scelfo, Berti, Silvestri  \& Viel}{Scelfo et~al.}{2023}]{Scelfo_2022_2}
Scelfo G.,  Berti M.,  Silvestri A.,   Viel M.,  2023, \mn@doi [Journal of Cosmology and Astroparticle Physics] {10.1088/1475-7516/2023/02/010}, 2023, 010

\bibitem[\protect\citeauthoryear{Schlegel et~al.,}{Schlegel et~al.}{2019}]{Astro2020}
Schlegel D.~J.,  et~al., 2019, Astro2020 APC White Paper: The MegaMapper: a z > 2 spectroscopic instrument for the study of Inflation and Dark Energy (\mn@eprint {arXiv} {1907.11171}), \url {https://arxiv.org/abs/1907.11171}

\bibitem[\protect\citeauthoryear{Schlegel et~al.,}{Schlegel et~al.}{2022}]{Megamapper}
Schlegel D.~J.,  et~al., 2022, The MegaMapper: A Stage-5 Spectroscopic Instrument Concept for the Study of Inflation and Dark Energy (\mn@eprint {arXiv} {2209.04322}), \url {https://arxiv.org/abs/2209.04322}

\bibitem[\protect\citeauthoryear{Talbot \& Thrane}{Talbot \& Thrane}{2018}]{Talbot_2018}
Talbot C.,  Thrane E.,  2018, \mn@doi [The Astrophysical Journal] {10.3847/1538-4357/aab34c}, 856, 173

\bibitem[\protect\citeauthoryear{Tansella, Bonvin, Durrer, Ghosh  \& Sellentin}{Tansella et~al.}{2018}]{Tansella_2018}
Tansella V.,  Bonvin C.,  Durrer R.,  Ghosh B.,   Sellentin E.,  2018, \mn@doi [Journal of Cosmology and Astroparticle Physics] {10.1088/1475-7516/2018/03/019}, 2018, 019

\bibitem[\protect\citeauthoryear{{The DESI Collaboration} et~al.,}{{The DESI Collaboration} et~al.}{2016}]{DESIp1}
{The DESI Collaboration} et~al., 2016, The DESI Experiment Part I: Science,Targeting, and Survey Design (\mn@eprint {arXiv} {1611.00036}), \url {https://arxiv.org/abs/1611.00036}

\bibitem[\protect\citeauthoryear{{Viljoen}, {Fonseca}  \& {Maartens}}{{Viljoen} et~al.}{2020}]{2020JCAP...09..054V}
{Viljoen} J.-A.,  {Fonseca} J.,   {Maartens} R.,  2020, \mn@doi [\jcap] {10.1088/1475-7516/2020/09/054}, \href {https://ui.adsabs.harvard.edu/abs/2020JCAP...09..054V} {2020, 054}

\bibitem[\protect\citeauthoryear{Viljoen, Fonseca  \& Maartens}{Viljoen et~al.}{2021}]{Viljoen_2021}
Viljoen J.-A.,  Fonseca J.,   Maartens R.,  2021, \mn@doi [Journal of Cosmology and Astroparticle Physics] {10.1088/1475-7516/2021/11/010}, 2021, 010

\bibitem[\protect\citeauthoryear{Wilson \& White}{Wilson \& White}{2019}]{Wilson_2019}
Wilson M.,  White M.,  2019, \mn@doi [Journal of Cosmology and Astroparticle Physics] {10.1088/1475-7516/2019/10/015}, 2019, 015–015

\bibitem[\protect\citeauthoryear{Ye \& Fishbach}{Ye \& Fishbach}{2021}]{YeFishback}
Ye C.,  Fishbach M.,  2021, \mn@doi [Physical Review D] {10.1103/physrevd.104.043507}, 104

\bibitem[\protect\citeauthoryear{Yoo, Fitzpatrick  \& Zaldarriaga}{Yoo et~al.}{2009}]{Yoo_2009}
Yoo J.,  Fitzpatrick A.~L.,   Zaldarriaga M.,  2009, \mn@doi [Physical Review D] {10.1103/physrevd.80.083514}, 80

\bibitem[\protect\citeauthoryear{Zazzera, Fonseca, Baker  \& Clarkson}{Zazzera et~al.}{2024}]{Zazzera_2024}
Zazzera S.,  Fonseca J.,  Baker T.,   Clarkson C.,  2024, \mn@doi [Journal of Cosmology and Astroparticle Physics] {10.1088/1475-7516/2024/05/095}, 2024, 095

\makeatother
\end{thebibliography}

\end{document}